\documentclass[letterpaper, 11pt]{article}
\usepackage[T1]{fontenc}

\usepackage{amsmath,amsfonts,amsthm,mathtools,color}
\usepackage{mathtools,color}
\usepackage{fullpage}
\usepackage{thmtools,thm-restate}
\usepackage[algo2e, linesnumbered,noend,ruled,noline]{algorithm2e}

\SetCommentSty{mycommfont}
\usepackage{setspace}
\usepackage{tikz}
\renewcommand{\algorithmcfname}{MECHANISM}
\SetAlFnt{\small}
\SetAlCapFnt{\small}
\SetAlCapNameFnt{\small}
\SetAlCapHSkip{0pt}
\IncMargin{-\parindent}

\usepackage{graphicx}
\usepackage{amsthm}
\usepackage[hidelinks]{hyperref}
\usepackage{cleveref}
\usepackage[numbers]{natbib}
\setcitestyle{acmnumeric}
\DontPrintSemicolon
\SetCommentSty{mycommfont}
\SetArgSty{textnormal}
\usepackage{soul}
\usepackage[title]{appendix}
\usepackage{apxproof}
\usepackage{enumitem}
\usepackage{verbatim}
\usepackage{authblk}

\newtheorem{theorem}{Theorem}
\newtheorem{lemma}{Lemma}
\newtheorem*{lemma*}{Lemma}
\newtheorem*{theorem*}{Theorem}
\newtheorem{corollary}{Corollary}
\newtheorem{claim}{Claim}

\newtheorem{observation}{Observation}

\title{Clock Auctions Augmented with Unreliable Advice
}

\author[a]{Vasilis Gkatzelis\thanks{Partially supported by NSF CAREER award CCF-2047907 and NSF grant CCF-2210502, CCF-2008280 and CCF-1755955.}}
\author[b]{Daniel Schoepflin\thanks{Partially supported by NSF grants CCF-2008280 and CCF-1755955 and by a grant to DIMACS from the Simons Foundation (820931). A portion of this work was supported by the NSF under Grant No. DMS-1928930
and by the Alfred P. Sloan Foundation under grant G-2021-16778, while this author was in residence at
the Simons Laufer Mathematical Sciences Institute (formerly MSRI) in Berkeley, California, during the Fall 2023
semester.}}
\author[a]{Xizhi Tan\thanks{Partially supported by NSF CAREER award CCF-2047907 and NSF grant CCF-2210502.}}
\affil[a]{Drexel University: \texttt{{\{gkatz,xizhi\}@drexel.edu}}}
\affil[b]{Rutgers University -- DIMACS: \texttt{ds2196@dimacs.rutgers.edu}}
\date{} 
\newcommand{\mech}{\mathcal{M}}
\newcommand{\opt}{\texttt{OPT}}

\newcommand{\feasible}{\mathcal{F}}
\newcommand{\pred}{\hat{\opt}}

\newcommand{\tol}{\overline{\eta}}
\newcommand{\consistency}{\alpha}
\newcommand{\robustness}{\beta}

\newcommand{\sconsistency}{consistency$^\infty$}

\newcommand{\uniform}{\textsc{UniformPrice}}
\newcommand{\rev}{\text{rev}}
\newcommand{\bidders}{N}

\newcommand{\increment}{\delta}

\newcommand{\revP}{R^P}
\newcommand{\price}{\mathbf{p}}
\newcommand{\errmech}{\textsc{ErrorTolerant}}
\newcommand{\cmech}{\textsc{FollowTheUnpredictedLeader}}

\begin{document}

\begin{titlepage}

\maketitle
\begin{abstract}
We provide the first analysis of (deferred acceptance) clock auctions in the learning-augmented framework. These auctions satisfy a unique list of very appealing properties, including obvious strategyproofness, transparency, and unconditional winner privacy, making them particularly well-suited for real-world applications. However, early work that evaluated their performance from a worst-case analysis perspective concluded that no deterministic clock auction with $n$ bidders can achieve a $O(\log^{1-\epsilon} n)$ approximation of the optimal social welfare for a constant $\epsilon>0$, even in very simple settings. This overly pessimistic impossibility result heavily depends on the assumption that the designer has no information regarding the bidders' values. Leveraging the learning-augmented framework, we instead consider a designer equipped with some (machine-learned) advice regarding the optimal solution; this advice can provide useful guidance if accurate, but it may be  unreliable.


Our main results are learning-augmented clock auctions that use this advice to achieve much stronger performance guarantees whenever the advice is accurate (known as \emph{consistency}), while maintaining worst-case guarantees even if this advice is arbitrarily inaccurate (known as \emph{robustness}). Our first clock auction achieves the best of both worlds: $(1+\epsilon)$-consistency for any desired constant $\epsilon>0$ and $O(\log{n})$ robustness; we also extend this auction to achieve error tolerance. We then consider a much stronger notion of consistency, which we refer to as consistency$^\infty$, and provide an auction that achieves a near-optimal trade-off between consistency$^\infty$ and robustness. Finally, using our impossibility results regarding this trade-off, we prove lower bounds on the ``cost of smoothness,'' i.e., on the robustness that is achievable if we also require that the performance of the auction degrades smoothly as a function of the prediction error. 

  \end{abstract}

\thispagestyle{empty} 
\end{titlepage}

\section{Introduction}\label{sec:intro}
Our focus in this paper is on a classic family of mechanism design problems involving a set \( N \) of \( n \) buyers competing for some type of resource or service, and a seller (the service provider) who needs to decide which subset of buyers to serve. Each buyer \( i \in N \) has a value \( v_i \) for receiving the service (i.e., the price they are willing to pay for it), and there is a feasibility constraint \( \feasible \subseteq 2^N \) that restricts the sets of agents that can be served simultaneously. The goal of the service provider is to achieve an efficient outcome, i.e., to serve a feasible set of buyers \( F \in \feasible \) that maximizes social welfare: \( \sum_{i \in F} v_i \). Depending on the nature of the feasibility constraint \( \feasible \), this captures many well-studied optimization problems, like the knapsack problem, the maximum weight independent set problem, and many more. However, apart from the computational challenges of the underlying optimization problem, the seller also faces the additional obstacle that the values of the buyers are private (not known to the seller), making it even harder to maximize social welfare.

To address the fact that buyers' values are private, a common solution in the mechanism design literature is to use \emph{direct revelation mechanisms}: these mechanisms ask each buyer to report their value to the mechanism, which can then use this information to decide which buyers should receive the service and how much they should pay for it. However, unless these mechanisms are carefully designed, they are very likely to introduce incentives for buyers to misreport their true values (e.g., to report a lower value to reduce their payment for the service). A central goal in mechanism design is to design \emph{strategyproof} mechanisms, which ensure that buyers' optimal strategy is to always report their true values. A classic example of a strategyproof direct revelation mechanism that maximizes social welfare is the Vickrey-Clarke-Groves (VCG) mechanism. However, despite its elegance and great theoretical appeal, the VCG mechanism is rarely used in practice due to a variety of shortcomings \cite{LonelyVCG}. For example, its strategyproofness heavily depends on its ability to solve the underlying optimization problem optimally, even though it may be NP-hard. In fact, direct revelation mechanisms in general face several issues that make them impractical for many applications. For example, (i) even if they are actually strategyproof, it may be non-trivial for buyers to verify it \cite{KHL1987}, (ii) they require that buyers put a lot of trust in the designer \cite{akbarpour2020credible}, and (iii) buyers need to directly reveal all of their private information, which can be sensitive and could potentially be used against them in the future (e.g., to set higher reserve prices) \cite{rothkopf1990vickrey}.

To overcome the issues that direct revelation mechanisms face, Milgrom and Segal \cite{MS2014,MS2019} proposed \emph{deferred-acceptance clock auctions} as a much more practical alternative for high-stakes auctions. In a clock auction, buyers are not asked to report their private values. Instead, the auction takes place over a sequence of rounds, and each buyer is offered a personalized price that weakly increases over time. Buyers can then remain active as long as they are willing to pay the price offered to them, and they can drop out whenever the price offered exceeds their value. These auctions have a wide variety of benefits that make them very practical: they guarantee (i) transparency (there is no way the auctioneer can mishandle the information provided by the buyers behind the scenes), (ii) unconditional winner privacy (the winners of the auction never need to reveal their true value), and (iii) simplicity (buyers do not need to understand the inner workings of the auction; all they need to know is the price offered to them in each round). These are properties that most strategyproof auctions do not satisfy.

Motivated by this unique list of very appealing properties, subsequent work focused on analyzing the worst-case approximation guarantees that clock auctions can achieve for a variety of different feasibility constraints, often setting aside computational constraints and focusing on information-theoretic constraints. The first set of results was rather pessimistic, showing that no deterministic clock auction can guarantee an approximation much better than \( O(\log n) \) even in seemingly simple settings \cite{DGR17}. On the positive side, subsequent work showed that there exists a natural clock auction, the water-filling clock auction (WFCA), that can guarantee an approximation of \( O(\log n) \) for any type of feasibility constraint \cite{CGS22}. Although the water-filling auction achieves the optimal worst-case approximation guarantee, a logarithmic approximation may not be quite as appealing from a practical standpoint. This is a common issue when analyzing algorithms and mechanisms from a worst-case perspective, leading to results that may be overly pessimistic.

To overcome the overly pessimistic nature of worst-case analysis, a surge of recent work has used the \emph{learning-augmented framework}, which assumes that the designer is provided with some useful, though unreliable, prediction or advice regarding the instance at hand. Using this advice, the goal is to provide stronger and much more practical performance guarantees whenever this advice is accurate (the \emph{consistency} guarantee), while simultaneously maintaining non-trivial worst-case guarantees even if the advice is arbitrarily inaccurate (the \emph{robustness} guarantee). As a result, this framework provides a natural way to leverage machine-learned predictions to guide the design of mechanisms while maintaining the important robustness that comes with worst-case analysis.
\subsection{Our Results}\label{sec:ourresult}
We provide the first analysis of clock auctions within the learning-augmented framework. Specifically, rather than assuming that the auction has absolutely no information regarding the values of the bidders, we assume it is provided with a prediction, \( \pred \in \feasible \), regarding which feasible set in \( \feasible \) has the optimal social welfare. Crucially, this prediction can be arbitrarily inaccurate, so the auction must use it carefully to maintain any bounded robustness.

In stark contrast to prior work, our first main result is highly optimistic: we propose a learning-augmented clock auction that simultaneously achieves the "best of both worlds." Specifically, our auction achieves a constant approximation of the optimal social welfare whenever the prediction is accurate, i.e., \( \pred \) is indeed optimal, while maintaining the worst-case approximation guarantee of \( O(\log n) \), regardless of how inaccurate the prediction may be. In fact, we show that it achieves a consistency of \( 1+\epsilon \) for any desired, arbitrarily small constant \( \epsilon > 0 \), while maintaining a robustness of \( O(\frac{1}{\epsilon} \log n) \), and we prove that this is asymptotically optimal even with respect to its dependence on \( \epsilon \). We also extend this auction to achieve error tolerance, i.e., an approximation guarantee that degrades gracefully as a function of the error, up to some predetermined error tolerance threshold.

For our second main result, we move beyond the standard notion of consistency and consider a significantly stronger notion, which we refer to as \sconsistency. In contrast to the standard consistency constraint that binds only if the prediction is fully accurate, \sconsistency\ binds on every input and requires that the social welfare achieved by the auction always approximates the social welfare of the predicted set, \( \pred \). Note that, whenever \( \pred \) is indeed optimal, approximating its social welfare reduces to approximating the optimal social welfare, so \sconsistency\ is strictly more demanding than the standard notion of consistency.

This notion of consistency dates back to earlier work by Mahdian et al. \cite{MNS07} and has also been studied explicitly or implicitly in some subsequent work in the learning-augmented literature. Our main result for this notion is a learning-augmented clock auction that achieves a near-optimal trade-off between robustness and \sconsistency. Specifically, we show that the optimal robustness achievable by any learning-augmented clock auction that satisfies \( \alpha \)-\sconsistency\ is \( \Omega(\frac{1}{\alpha}n^{1/(\consistency-1)} \log{n}) \) and our auction combines \( \alpha \)-\sconsistency\ with $O(n^{1/(\consistency-1)} \log{n})$ robustness. Note that for any constant $\alpha$ \sconsistency, the robustness of our auction is asymptotically optimal.

Finally, using our robustness lower bound for constant \sconsistency, we also prove a lower bound on the ``cost of smoothness'' for clock auctions, i.e., a lower bound on the best achievable robustness if we also require that the performance of the auction degrades smoothly as a function of the prediction error. This result exhibits an interesting connection between \sconsistency\ and smoothness and is in contrast with almost all prior work in the learning-augmented framework, which does not provide impossibility results regarding smoothness guarantees.

\subsection{Related Work}\label{sec:relatedwork}
Our paper lies at the intersection of two lines of literature: the literature on (deferred acceptance) clock auctions and the literature on learning-augmented algorithm and mechanism design. Below, we provide an overview of the most relevant papers in each of these two areas.

\paragraph{Clock auctions.}
Since their formal introduction in \cite{MS2014,MS2019}, clock auctions have attracted attention from both the economics and computer science communities, partly due to their many practical advantages over the sealed-bid format. These important practical properties include: (i) obvious strategyproofness, a strong notion of incentive compatibility originally defined by Li \cite{li2017obviously}; (ii) unconditional winner privacy, defined by Milgrom and Segal \cite{MS2019}, which guarantees that clock auctions reveal the minimum possible information about the winners' values; and (iii) transparency, including the fact that ascending clock auctions are  ``credible'' auctions, as defined in \cite{akbarpour2020credible}. Notably, clock auctions are the \emph{unique} class of auctions satisfying these properties, making them particularly well-suited to practical implementation. We refer the reader to \cite{MS2019} and \cite[Section 1.2]{FGGS22} for a more complete discussion of the properties of clock auctions.

The papers technically closest to our work include \cite{BLP2009, CGS22, DGR17}. Before \cite{MS2014, MS2019} explicitly defined clock auctions, Babaioff et al. \cite{BLP2009} proposed a deterministic, prior-free clock auction that achieved an \( O(\log{v_{\max}}) \)-approximation, where \( v_{\max} \) is the highest value of any single bidder. However, their auction notably achieved only a \( \Theta(n) \)-approximation. In terms of lower bounds, Dütting et al. \cite{DGR17} demonstrated that no deterministic, prior-free clock auction could achieve a \( \log^{\tau}{n} \)-approximation for any constant \( \tau < 1 \). Over a decade after \cite{BLP2009}, Christodoulou et al. \cite{CGS22} showed that the bound in \cite{DGR17} was essentially tight by providing a deterministic, prior-free clock auction that achieved an \( O(\log{n}) \)-approximation (and simultaneously an \( O(\log{v_{\max}}) \)-approximation).

In single-parameter, forward auction settings, Christodoulou et al. \cite{CGS22} were the first to propose using \emph{randomization} to circumvent the impossibility results of \cite{DGR17}. They presented an auction that guarantees \emph{expected} welfare, achieving an \( O(\sqrt{\log{k}}) \)-approximation to the optimal social welfare, where \( k \) is the number of maximal feasible sets in the set system \( \mathcal{F} \). This was later improved to an \( O(\log{\log{k}}) \)-approximation guarantee by \cite{FGGS22}.

Clock auctions have also been studied beyond the single-parameter, binary service level, forward auction context. For example, \cite{GMR17} demonstrated how to extend clock auctions to settings with multiple levels of service and analyzed several special cases of this general framework. Many works have also examined \emph{reverse} (i.e., procurement) clock auctions, both with \cite{JM17, BGGST22, ensthaler2014dynamic, BB2023, HHCT23} and without \cite{K2015, bichler2020strategyproof} an auctioneer budget constraint. The works of \cite{LM2020, DTR2017} studied double clock auctions involving both strategic sellers and buyers. The work of \cite{GPPS21} studied clock auctions in the economically significant setting of interdependent values. Finally, the work of \cite{GH23} examined clock auctions aimed at maximizing consumer utility.

\paragraph{Learning-augmented framework.}
In recent years, the learning-augmented framework has gained widespread recognition as a valuable paradigm for designing and analyzing algorithms. We refer to \cite{MV22} for a survey of early contributions and \cite{alps} for a frequently updated list of papers in this field. This approach aims to overcome the limitations of overly pessimistic worst-case bounds. In the past five years alone, over 200 papers have revisited classic algorithmic problems using this framework, with prominent examples including online paging \cite{lykouris2018competitive}, scheduling \cite{PSK18}, optimization problems involving covering \cite{BMS20} and knapsack constraints \cite{IKQP21}, as well as Nash social welfare maximization \cite{banerjee2020online}, the secretary problem \cite{AGKK23, DLLV21, KY23}, and a variety of graph-related challenges \cite{azar2022online}. As we discuss in our results section, there is another, stronger notion of consistency that requires a good approximation of the \emph{predicted solution} irrespective of the quality of predictions. This notion was first proposed in \cite{MNS07}, and subsequent work, including \cite{BMS20, KN23}, also defines it as consistency. Additionally, we note that while some works do not explicitly state it, their results actually hold for this stronger notion of consistency \cite{WZ20, JLLTZ22, ABGOT22}.

More closely related to our work, the line of research on learning-augmented mechanisms interacting with strategic agents is recent and was initiated by Agrawal et al. \cite{ABGOT22} and Xu and Lu \cite{XL22}. This line of work encompasses strategic facility location \cite{ABGOT22, XL22, IB22, BGT24, chen2024strategic, balkanski2024randomized}, strategic scheduling \cite{XL22, BGT223}, auctions \cite{MV17, XL22, LuWanZhang23, caragiannis2024randomized, BGTZ23}, bicriteria mechanism design (optimizing both social welfare and revenue) \cite{BPS23}, graph problems with private input \cite{CKST24}, metric distortion \cite{BFGT23}, and equilibrium analysis \cite{GKST22, IBB24}. Recently, \cite{CSV24} revisited mechanism design problems with predictions on the outcome space instead of the input. The prediction discussed in our work is one such example. For further discussion regarding this line of work, we refer the reader to \cite{BGT23}.

\section{Preliminaries}\label{sec:prelim}
We consider a canonical single-parameter auction setting where an auctioneer aims to allocate a service among a set \( \bidders \) of \( n \) bidders. There exists a public family of sets \( \feasible \subseteq 2^\bidders \) indicating subsets of bidders that can feasibly be served by the auctioneer. We assume \( \feasible \) is \emph{downward-closed}, meaning if \( S \in \feasible \), then \( T \subseteq S \) implies \( T \in \feasible \). The auctioneer's goal is to select a feasible subset \( S \in \feasible \) of bidders to maximize social welfare, i.e., \( v(S) = \sum_{i \in S} v_i \). Since bidders have private values, the auctioneer must design a mechanism to elicit these values. However, bidders are selfish and may strategically misreport their values to achieve their preferred outcomes (i.e., receiving the service at a lower price). Therefore, the auctioneer must carefully design payment rules to incentivize accurate reporting.

We design clock auctions for this problem. A clock auction is a dynamic mechanism proceeding over several rounds. In each round \( t \), each bidder \( i \) is offered a personalized ``clock'' price \( p_{i,t} \), which represents the current amount they would need to pay if the auction were to terminate at that moment. At the outset of the auction, the clock price \( p_{i,0} \) for each bidder \( i \) is initialized to some arbitrarily small positive value \( v_{\min} \), and each bidder is placed in an ``active'' bidder set \( A \). The clock prices are non-decreasing throughout the auction, i.e., \( p_{i, t} \geq p_{i, t-1} \) for all rounds \( t \).\footnote{We assume that each bidder with positive value has a value above a publicly known minimum value \( v_{\min} \), which is consistent with \cite{BLP2009} and \cite{CGS22}. Thus, rather than continuous prices, we may increment prices by an arbitrarily small positive value \( \increment \), say, \( v_{\min}/n^2 \). As argued in \cite{CGS22}, discretizing prices to this range causes negligible loss in approximation since the optimal social welfare is lower-bounded by \( v_{\min} \).} In each round \( t \), the auctioneer announces a price \( p_{i,t} \) for each bidder \( i \in A \), and bidder \( i \) can then choose to remain in the auction (staying “active”) or permanently exit (becoming “inactive” and being removed from \( A \)). If \( i \) exits at any point in the auction, they will not receive the service and pay nothing. The auction then terminates when \( A \in \feasible \), i.e., all active bidders may be feasibly served.

Note that in each round, the clock price for each bidder can only depend on public information, i.e., the feasibility constraint, the history of prices, and the points at which bidders exited the auction. We use \( \rev(S,\mathbf{p_t}) = \sum_{i \in S \cap A}{p_{i,t}} \) to denote the \emph{revenue} of a set \( S \) at the current price vector \( \mathbf{p_t} \). In other words, the revenue of a set \( S \) in round \( t \) of the auction is the sum of the clock prices in round \( t \) of the active bidders in \( S \).

To analyze the performance of a clock auction \( \mech \), we consider its worst-case approximation guarantee. On an instance \( I \), let \( I \) denote the welfare obtained by the clock auction on \( I \) and \( \opt(I) \) denote the maximum social welfare among all feasible sets in \( I \). Then, \( \mech \) obtains an \( \alpha \)-approximation of the optimal social welfare for a class of instances \( \mathcal{I} \) if
\[
\max_{I \in \mathcal{I}} \frac{v(\opt(I))}{v(\mech(I))} \leq \alpha.
\]

\paragraph{The Water-Filling Clock Auction.}

The central result of \cite{CGS22} is a deterministic clock auction called the “Water-Filling Clock Auction” (WFCA), which achieves the (near) optimal approximation guarantee of \( O(\log{n}) \) for any family of instances where \( \feasible \) is downward-closed. The WFCA computes the revenue of each set in each round and marks the set of bidders \( W \) with the highest revenue as the “conditional winners” and increments the price of the lowest-priced bidders outside \( W \). This, intuitively, makes sense since the revenue of a set is a lower bound on its remaining possible welfare. As the WFCA is an important subroutine in our auctions, we include pseudocode for it here as Subroutine \ref{alg:wfca} for completion.

\renewcommand*{\algorithmcfname}{SUBROUTINE}
\begin{algorithm2e}[ht] \label{alg:wfca}
\SetKwInOut{Input}{Input}
\Input{An initial price vector $\price$ and set of bidders $\tilde{\bidders}$}

$t\leftarrow 0$, $A \leftarrow \tilde{\bidders}$, and $p_{i,t} \leftarrow \tilde{p}_i$ for all $i \in \tilde{\bidders}$

\While{$A \notin \feasible$}{
    $t\leftarrow t+1$
    
    $W \leftarrow \arg\max_{S \in \feasible: S \subseteq A}\left\{\sum_{i \in S}{p_{i,t-1}}\right\}$ be the set of highest current revenue \label{alg:line:W}
    
    $\ell \leftarrow \min_{i \in A \setminus W}\{p_{i,t-1}\}$ be the lowest price among active bidders not in \( W \) \label{alg:line:p}

    \ForEach{bidder $i \in A \setminus W$ with $p_{i,t-1}=\ell$}{
        $p_{i,t} \leftarrow p_{i,t-1} + \increment$ \tcp*{Increment prices of lowest-priced ``losers''}
    
        \If{$i$ rejects updated price}
        {
            $A \leftarrow A \setminus \{ i \}$
        }
    }

    \ForEach{bidder $i \in W$ and each bidder with $p_{i,t-1} > \ell$}{
        $p_{i,t} \leftarrow p_{i,t-1}$ \tcp*{Keep all other prices the same}
    }
}
\Return{A, $\price$} 
 \caption{The \emph{Water-Filling Clock Auction} (WFCA) of \cite{CGS22}}
\end{algorithm2e}

We summarize two critical features of the WFCA shown in \cite{CGS22} that we use in our analysis as Lemma \ref{lem:WFCA-rev-monotone} below. We note that the authors of \cite{CGS22} show that the WFCA achieves a \( 4\log{n} \)-approximation \cite[Theorem 2]{CGS22}. The proof of this theorem demonstrates a slightly stronger guarantee that is more useful for our bounds, namely that the welfare achieved is a \( 2H_n \)-approximation, where \( H_n = \sum_{i =1}^{n}{1/i} \) is the \( n \)-th harmonic number. It is known that \( H_n \) is asymptotically \( O(\log{n}) \).

\begin{lemma}[\cite{CGS22}]\label{lem:WFCA-rev-monotone}
    The Water-Filling Clock Auction is \emph{revenue monotone}, i.e., the sum of prices of bidders in the set with the highest sum of prices is monotonically increasing throughout the WFCA. In addition, the WFCA obtains a \( 2H_n \)-approximation to the optimal social welfare in any downward-closed set system.
\end{lemma}

\paragraph{The Learning-Augmented Framework.}

In this work, we adopt the learning-augmented framework and study clock auctions that are also equipped with a (potentially erroneous) prediction \( \pred \) regarding the feasible set of highest welfare. Given the predictions and an instance, we denote the welfare achieved by a mechanism \( \mech \) as \( \mech(I, \pred) \), and we evaluate the performance of \( \mech \) using its \emph{robustness} and \emph{consistency}.\footnote{We note that the predictions used by mechanisms are \emph{public}, i.e., the mechanism designer and bidders observe the predictions. Although our clock auctions utilize the predictions to guide the price-increase process, the bidders still face only monotonically increasing clock prices and, thus, they still have a simple interface with an obviously dominant strategy of exiting the auction when the clock price becomes undesirable (higher than their private value).}

The robustness of a mechanism refers to the worst-case approximation ratio of the mechanism given an adversarially chosen, possibly erroneous, prediction. Mathematically,
\[
\text{robustness}(\mech) = \max_{I, \pred} \frac{v(\opt(I))}{\mech(I, \pred)}.
\]

The consistency of a mechanism refers to the worst-case approximation ratio of the mechanism when the prediction it is provided with is accurate, i.e., \( \pred = \opt \). Mathematically, 
\begin{align}\label{def:consistency1}
    \text{consistency}(\mech) = \max_{I}\frac{v(\opt(I))}{\mech(I, \opt(I))}.
\end{align}

Additionally, we consider a stronger notion of consistency first proposed by \cite{MNS07}, which we denote as \sconsistency. It is defined to be the worst-case approximation ratio with respect to \emph{the value of the predicted set}, regardless of the correctness of the prediction. Mathematically, 
\begin{align}\label{def:consistency2}
    \text{\sconsistency}(\mech) = \max_{I, \pred} \frac{v(\pred)}{\mech(I, \pred)}.
\end{align}
We emphasize that an \( \alpha \)-\sconsistency\ guarantee implies an \( \alpha \)-consistency guarantee for any \( \alpha \) by considering a subset of instances where \( \pred = \opt(I) \).

\section{Clock Auctions Achieving the Best of Both Worlds}\label{sec:bobw}
In this section, we present a deterministic clock auction augmented with a prediction \( \pred \in \feasible \) that suggests the feasible set which maximizes social welfare. Our auction leverages this prediction to achieve a favorable trade-off between consistency (see Equation~\eqref{def:consistency1}) and robustness. Recall that without predictions, no mechanism can achieve an \( O(\log^{\tau} n) \)-approximation for any constant \( \tau < 1 \). The aim is to bypass this logarithmic barrier whenever the prediction is accurate, without significantly sacrificing robustness. 

One naive approach is to simply serve the predicted set, which would yield 1-consistency; however, this would suffer from unbounded robustness. Conversely, ignoring the prediction and simply running the WFCA would provide \( O(\log n) \) consistency and \( O(\log n) \) robustness. A desirable outcome is small constant consistency with \( O(\log n) \) robustness, often referred to as a “best-of-both-worlds” guarantee. This is precisely what our clock auction, formally defined in Mechanism~\ref{alg:follow-the-leader}, achieves. In fact, we can set the consistency arbitrarily close to \(1+\epsilon\) for any small constant \( \epsilon > 0 \), while maintaining \( O\left(\frac{1}{\epsilon} \log n\right) \)-robustness. We complement this positive result with a lower bound, showing that this robustness bound is tight with respect to its dependence on both \( n \) and \( \epsilon \).

\begin{theorem}\label{thm:bobw}
    For any downward-closed set system, Mechanism~\ref{alg:follow-the-leader} with parameter \( \gamma = \frac{10(1+\epsilon)}{9\epsilon} \) is \( (1+\epsilon) \)-consistent and \( O\left(\frac{1}{\epsilon}\log{n}\right) \)-robust for any constant \( \epsilon > 0 \).
\end{theorem}

The mechanism, called \cmech, is a clock auction that alternates between raising the price for bidders in the predicted and unpredicted sets, while maintaining a balanced ratio between the target revenue from the predicted set and the rejected welfare of the unpredicted set. At a high level, in each iteration, the mechanism first sets a new, higher revenue ``target'' \( R_t \) and pushes the unpredicted bidders only until the robustness guarantee based on current revenue is nearly at risk. It then further pushes the unpredicted bidders, aiming to ensure the robustness guarantee is satisfied solely by the unpredicted set. If all unpredicted bidders reject the offers during this process, the mechanism outputs the remaining active bidders in the predicted set. Otherwise, it switches to pushing the predicted set to match the current revenue obtained. If the target revenue is met, the process repeats; if all predicted bidders reject, the WFCA is run on the remaining unpredicted bidders. We assume that bidders choose the obviously dominant strategy of exiting the auction whenever the clock price exceeds their value, and we learn each bidder’s value upon their exit.

For a formal description, please refer to Mechanism~\ref{alg:follow-the-leader}.

\renewcommand*{\algorithmcfname}{MECHANISM}
\begin{algorithm2e}[h]
\DontPrintSemicolon
\LinesNumbered
\SetNoFillComment
\KwIn{A system $\feasible$ of feasible sets of bidders, a predicted optimal set $\pred$, a parameter $\gamma$}

$F \gets F \setminus \{F \cap \pred\}$ for all $F \neq \pred \in \feasible$ \tcp*{Make $\pred$ disjoint from all unpredicted sets}
\label{alg:line:make-disjointc}

$p_i \gets v_{\text{min}}$ for all $i \in N$ \tcp*{Raise price of each bidder to the initial small amount}
$R_0 \gets \rev(\pred, \mathbf{p})$

$t \leftarrow 1$ \tcp*{Initialize the counter}

\While{TRUE}{
$R_t \gets 10\cdot R_{t-1}$

Run $\uniform(N \setminus \pred, \price)$ until $\max_Fv(F \setminus N) \in [R_t\gamma H_n,2R_t\gamma H_n)$
\tcp*{Reject ``safe'' amount of welfare to determine minimum total value in optimal set\footnotemark}\label{alg:line:Uwelc}

\lIf{$N = \pred$}{
    \Return{$\pred$, $\price$ \label{alg:line:UwelTermc}}
}

Run $\uniform(N \setminus \pred, \price)$ until $\max_F\rev(F\cap N, \price) \geq R_t$ \tcp*{Ensure unpredicted sets ``cover'' their own lost welfare}\label{alg:line:Urevc}

\lIf{$N = \pred$}{
    \Return{$\pred$, $\price$ \label{alg:line:UrevTermc}}
}
Run $\uniform(\pred, \price)$ until $\rev(\pred, \price) \geq R_t$ \label{alg:line:Prevc} \tcp*{Ensure predicted set ``covers'' welfare lost from unpredicted sets}

\If{$\pred = \emptyset$}{
    \Return{ WFCA$(N, \price)$}\label{alg:line:WFCAc}
}

$t \leftarrow t + 1$
}
\SetAlgoRefName{1}
\caption{\cmech}
\label{alg:follow-the-leader}
\end{algorithm2e}

\renewcommand*{\algorithmcfname}{SUBROUTINE}
\begin{algorithm2e}
\DontPrintSemicolon
\LinesNumbered
\SetNoFillComment
\KwIn{A set $S$ and the current price $p_i$ for all $i \in S$, the current price vector $\price$}
\While{}{
$\ell \gets \min_{i \in S} p_i$

\For{$i$ with $p_i = \ell$}{
    $p_i \gets p_i +\epsilon$
    
    \If{$i$ reject the updated price $i$}{
        $S \gets S \setminus \{i\}$
    }
}
}
\caption{\uniform}
\label{alg:uniform}
\end{algorithm2e}
\renewcommand*{\algorithmcfname}{Mechanism}
\footnotetext{We can make sure the total welfare in some unpredicted set is at least $R_t\gamma H_n$ but the rejected welfare is no more than $2R_t\gamma H_n$ by stopping when current price of active unpredicted bidders reaches $R_t\gamma H_n$.}
We first show that Line~\ref{alg:line:make-disjointc} affects neither consistency nor robustness by too much. 
\begin{observation}\label{lem:make-disjointc}
    The social welfare of the optimal set in $\feasible$ before making the transformation in line \ref{alg:line:make-disjointc} of Mechanism~\ref{alg:follow-the-leader} is at most $2$ times greater than the social welfare of the optimal set after the transformation. 
\end{observation}
\begin{proof}
    We first note that the social welfare of the predicted set is the same before and after completing the transformation in line \ref{alg:line:make-disjoint} of Mechanism~\ref{alg:follow-the-leader}. Then if the predicted set was the optimal set before the transformation, then the value of the optimal set does not decrease.  So suppose some non-predicted set $O^*$ is optimal and denote its total welfare $v(O^*)$.  If at least half of the welfare contained in $O^*$ comes from bidders contained in the predicted set then after the transformation the predicted set has welfare at least $v(O^*)/2$ and, hence, the optimal set has welfare at least $v(O^*)/2$ after the transformation.  On the other hand, if more than half of the welfare contained in $O^*$ comes from bidders not in the predicted set then after the transformation $O^*$ still has social welfare at least $v(O^*)/2$. 
\end{proof}

For ease of exposition, for the remainder of this section, we assume that \( \feasible \) is such that the predicted set is completely disjoint from all maximal unpredicted sets. According to Observation \ref{lem:make-disjointc}, the \emph{consistency} guarantees we obtain in this special case are exactly the same as those we obtain for general set systems. Moreover, by Observation \ref{lem:make-disjointc}, the \emph{robustness} guarantees we obtain in this special case are within a factor of 2 of the robustness guarantees we obtain for general set systems. Therefore, it suffices to show that we obtain \( 1+\epsilon \)-consistent and \( O(\log{n}) \)-robust algorithms for the special case where the predicted set is completely disjoint from all unpredicted sets. We devote the rest of this section to demonstrating just that. For ease of presentation, without loss of generality, we normalize \( R_0 \) to 1. All the missing proofs can be found in Appendix~\ref{app:bobw}.

\subsubsection*{Robustness Argument.}
We first establish the robustness of Mechanism~\ref{alg:follow-the-leader}. To begin, we demonstrate the robustness guarantee if a subset of the predicted set is output (Lemma~\ref{lem:Robust-Pred-winsc}). For cases where an unpredicted set is output, we argue that the revenue obtained at the time point where \( \pred = \emptyset \) provides a good approximation of the best welfare achievable from any rejected subset (Lemma~\ref{lem:welfare-before-WFCAc}). We then utilize the revenue monotonicity of WFCA to achieve the desired robustness bound (Lemma~\ref{lem:unpred-wins-robustc}).

\begin{lemma} \label{lem:welf-single-round-U2c}
    The total welfare rejected from any unpredicted set in line \ref{alg:line:Urev} in any iteration $t$ of while loop of Mechanism~\ref{alg:follow-the-leader} is at most $R_t\cdot H_n$.
\end{lemma}
\begin{lemma}\label{lem:Robust-Pred-winsc}
    For any $\gamma > 10/9$, if Mechanism~\ref{alg:follow-the-leader} outputs a subset of the predicted set, then it is $\frac{9}{100(2\gamma + 1)H_n}$-robust.
\end{lemma}
\begin{proof}
    Consider the iteration $\hat{t}$ in which the while loop of Mechanism~\ref{alg:follow-the-leader} terminates.  By definition, the mechanism terminates in either line \ref{alg:line:UwelTermc} or line \ref{alg:line:UrevTermc}. By Line~\ref{alg:line:Uwelc} and Lemma \ref{lem:welf-single-round-U2c} we have that the total welfare rejected from any unpredicted set in any round $t'$ is at most $2R_{t'}\gamma H_n + R_{t'}H_n = R_{t'} \cdot (2\gamma + 1)H_n$.  But then, summing over all rounds we have that the total welfare contained in any unpredicted set is at most \[\sum_{t' = 1}^{\hat{t}}{R_{t'} \cdot (2\gamma + 1)H_n} = (2\gamma + 1)H_n \cdot \sum_{t' = 1}^{\hat{t}}{10^{t'}} = (2\gamma + 1)H_n \cdot \frac{10^{\hat{t} + 1} - 1}{9} \leq R_{\hat{t}} \cdot \frac{(2\gamma + 1)10H_n}{9}.\]  On the other hand, since we continued to iteration $\hat{t}$ we have that the revenue of the bidders in the predicted set reached $R_{\hat{t} - 1} =\frac{R_{\hat{t}}}{10}$.  
    Hence, Mechanism~\ref{alg:follow-the-leader} obtains at least a $\frac{R_{\hat{t}}/10}{R_{\hat{t}} \cdot \frac{(2\gamma + 1)10H_n}{9}} = \frac{9}{ 100(2\gamma + 1)H_n}$-fraction of the optimal social welfare.
\end{proof}
\begin{lemma}\label{lem:welfare-before-WFCAc}
    For any $\gamma > 10/9$, if Mechanism~\ref{alg:follow-the-leader} reaches line \ref{alg:line:WFCA}, then the revenue of active bidders in the highest revenue set before running WFCA is at least a $\frac{9}{10(2\gamma + 1) H_n}$-approximation of the maximum social welfare obtainable by rejected unpredicted bidders, i.e., 
    \[\left(\frac{10(2\gamma + 1) H_n}{9}\right)\max_{F \in \feasible}\rev(F \cap N, \price) \geq \max_{F \in \feasible} v(F \setminus N).\]
\end{lemma}
\begin{proof}
    As before we consider the iteration $\hat{t}$ in which the while loop of Mechanism~\ref{alg:follow-the-leader} hit Line~\ref{alg:line:WFCAc} By assumption, our while loop terminates in line \ref{alg:line:WFCAc}. Following the same reasoning as Lemma \ref{lem:Robust-Pred-winsc}, the total welfare rejected from any unpredicted set over all rounds from $1$ to $\hat{t}$ is at most \[\sum_{i' = 1}^{\hat{t}}{R_{i'} \cdot (2\gamma + 1)H_n} \leq R_{\hat{t}} \cdot \frac{(2\gamma + 1)10H_n}{9}.\]  On the other hand, because we arrive at line \ref{alg:line:WFCAc} we know that there exists some set unpredicted set which obtains revenue $R_{\hat{t}}$ in round $\hat{t}$.  Thus, the revenue among active bidders in the highest revenue set when we reach line \ref{alg:line:WFCAc} is a $\frac{9}{10(2\gamma + 1) H_n}$-fraction of the maximum social welfare obtained from any feasible set of rejected unpredicted bidders.
\end{proof}

\begin{lemma}\label{lem:unpred-wins-robustc}
    For any $\gamma > 10/9$, if Mechanism~\ref{alg:follow-the-leader} outputs a subset of some unpredicted sets, then it is $\frac{9}{10(2\gamma + 1) H_n} + \frac{1}{2H_n}$-robust.
\end{lemma}
\begin{proof}
    In case Mechanism~\ref{alg:follow-the-leader} outputs a subset of unpredicted bidders it must arrive at line \ref{alg:line:WFCAc}.  As such, we can decompose the social welfare from the optimal (unpredicted) set into two components, the portion of the welfare coming from bidders rejected before Mechanism~\ref{alg:follow-the-leader} runs the Water-filling Clock Auction and the portion of the welfare coming from bidders who participate in the WFCA.  From Lemma \ref{lem:welfare-before-WFCAc} we know that the \emph{revenue} reached by Mechanism~\ref{alg:follow-the-leader} before running the WFCA is within a $\frac{9}{10(2\gamma + 1) H_n}$-factor of the welfare rejected from the optimal set before running the WFCA.  Moreover, from Lemma \ref{lem:WFCA-rev-monotone} we have that the revenue reached by Mechanism~\ref{alg:follow-the-leader} after running the WFCA is weakly higher than the revenue reached before running the WFCA.  As such, since the welfare obtained by serving a set of bidders is always weakly higher than the revenue collected from these bidders we have that the social welfare obtained by Mechanism~\ref{alg:follow-the-leader} is within a $\frac{9}{10(2\gamma + 1) H_n}$-factor of the welfare rejected from the optimal set before running the WFCA.  Finally, we have from Lemma \ref{lem:WFCA-rev-monotone} that the social welfare achieved by running the WFCA only on bidders who are active when line \ref{alg:line:WFCAc} is reached is within a $1/(2H_n)$-factor of the optimal social welfare achievable from  these bidders.  Combining these two guarantees, we have that the social welfare obtained by Mechanism~\ref{alg:follow-the-leader} is within a $\frac{9}{10(2\gamma + 1) H_n} + \frac{1}{2H_n}$-factor of the optimal social welfare.
\end{proof}

\subsubsection*{Consistency Argument.}
We now establish the consistency of Mechanism~\ref{alg:follow-the-leader}. The high-level idea is to upper bound the total rejected welfare when the mechanism terminates and argue that: (i) If the predicted set is indeed optimal, the mechanism will always output a subset of the predicted set; and (ii) the remaining welfare in the predicted set (when it is output) is high as a result of us ``favoring'' the predicted set in Mechanism \ref{alg:follow-the-leader}.

\begin{lemma}\label{lem:total-rejected-Predc}
    The total welfare rejected from the predicted set up to the end of the $t$-th iteration of the while loop is at most $R_t\cdot 10H_n/9$
\end{lemma}
\begin{lemma}\label{lem:ftl-consistentc}
    For any $\gamma > 10/9$, Mechanism~\ref{alg:follow-the-leader} is $\left(1 - \frac{10}{9\gamma}\right)$-consistent.
\end{lemma}
\begin{proof}
  Fix an instance $I$ where the optimal set (also the predicted set by definition of consistency) has social welfare $V^*$.
    Consider the minimum value $k$ such that $10^{k} \cdot \gamma H_n$ exceeds $V^*$.  Let $\hat{t}$ denote the value of $t$ at the beginning of the final iteration of the while loop of Mechanism~\ref{alg:follow-the-leader} when run on instance $I$. We first note that the Mechanism~\ref{alg:follow-the-leader} terminates in either Line \ref{alg:line:UwelTermc}
     or Line \ref{alg:line:UrevTerm}, otherwise there exist a unpredicted set with value more than $V^*$.  Since we only continue to line \ref{alg:line:Urevc} after the preceding if statement when there are still active unpredicted bidders it must be that $\hat{t} \leq k$ (as, otherwise, some unpredicted set would have total welfare greater than $V^*$) and $\hat{t}= k$ if we terminate in Line~\ref{alg:line:UwelTermc} (i.e., before we cause any additional predicted bidders to exit the auction in the iteration $k$).  
    
    Now consider the total amount of welfare lost from the predicted set from rounds $1$ to $\hat{t}-1$.  By Lemma \ref{lem:total-rejected-Predc} we have that this total welfare is at most $R_{\hat{t}-1} \cdot 10H_n/9 = 10^{\hat{t} - 1} \cdot \frac{10H_n}{9} \leq 10^{k-1} \cdot \frac{10H_n}{9}$.  On the other hand, by our definition of $k$, we know that the total social welfare in the predicted set is at least $10^{k-1} \cdot \gamma H_n$.  But then, at the end of the $\hat{t}-1$ iteration there must exist active predicted bidders and, moreover, the fraction of the predicted set's social welfare that we retain is at least 
    \[\frac{10^{k-1}\cdot \gamma H_n - 10^{k-1}\cdot \frac{10}{9}H_n}{10^{k-1}\cdot \gamma H_n} = \frac{\gamma - \frac{10}{9}}{\gamma},\]
    which completes the proof.
\end{proof}

We are now ready to prove the main theorem of the section.
\begin{proof}(\emph{Proof for Theorem~\ref{thm:bobw}})
    First observe that $\epsilon > 0$ ensures that $\gamma > 10/9$.  In the case that the predicted set, indeed, optimal, Lemma \ref{lem:ftl-consistentc} gives that some subset of the predicted set is served.  Moreover, we obtain that the total social welfare of served bidders is within a factor \[\left(1 - \frac{10}{9\gamma}\right) = \left(1 - \frac{10}{9\frac{10(1+\epsilon)}{9\epsilon}}\right) = \frac{1}{1+\epsilon}\] of the total social welfare in the predicted set, thereby guaranteeing $(1+\epsilon)$-consistency.  On the other hand, if some subset of the predicted set is served, Lemma \ref{lem:Robust-Pred-winsc} guarantees that the total social welfare obtained by Mechanism~\ref{alg:follow-the-leader} is within a factor $\frac{9}{100(2\gamma + 1)H_n}$ of the optimal social welfare.  Since $H_n = O(\log{n})$ and $\gamma$ is a constant for any choice of constant $\epsilon$ we obtain $O(\log{n})$-robustness.  Finally, if some subset of an unpredicted set is served, Lemma \ref{lem:unpred-wins-robustc} guarantees that Mechanism~\ref{alg:follow-the-leader} obtains welfare within a factor $\frac{9}{10(2\gamma + 1)H_n} + \frac{1}{2H_n}$ of the optimal welfare.  Again, since $H_n = O(\log{n})$ and $\gamma$ is constant for any constant $\epsilon$ we obtain $O(\log{n})$-robustness in this case.
\end{proof}

While the focus of our work is on leveraging advice to circumvent the informational limitations faced by clock auctions (recall that the lower bound from \cite{DGR17} precludes clock auctions from achieving a \( \log^{\tau}(n) \)-approximation for any constant \( \tau < 1 \)), \cmech\ could still require exponential time for certain feasibility constraints (e.g., when \( \feasible \) corresponds to the independent sets of a graph). Fortunately, a key advantage of \cmech\ is that it admits the use of a black-box approximation algorithm for the welfare and revenue maximization steps in lines 7 and 9. Specifically, if there exists an algorithm that outputs a set of rejected bidders which gives a \( \rho \)-approximation to the welfare of the best-rejected set in line 7 and an active set which gives \( \rho \)-approximation to the revenue in line 9, then our algorithm achieves robustness that scales linearly with \( \rho \) (i.e., robustness becomes \( O\left(\rho \cdot \frac{1}{\epsilon} \log n\right) \)). This result is formalized in Corollary~\ref{cor:approx-ftul}. For completeness, a full proof is provided in Appendix~\ref{app:bobw}.

\begin{corollary}\label{cor:approx-ftul}
    For any downward-closed set system, Mechanism \ref{alg:follow-the-leader} with parameter $\gamma = \frac{10(1+\epsilon)}{9\epsilon}$ equipped with a $\rho$-approximate algorithm for the underlying problem of maximizing revenue/welfare (at some fixed clock prices) is $(1+\epsilon)$-consistent and $O(\rho \cdot \frac{1}{\epsilon}\cdot \log{n})$-robust for any constant $\epsilon > 0$.
\end{corollary}

\subsection{Lower Bound}\label{sec:bobwlowerbound}
We now show that apart from achieving the ``best of both worlds'' in terms of asymptotic robustness and consistency guarantees, our learning-augmented clock auction also achieves a tight asymptotic dependence on the $\epsilon$ parameter. Specifically, we show that there exists a family of instances for which any $(1+\epsilon)$-consistent auction has robustness at least $\Omega(\frac{1}{\epsilon}\log n)$ (the proof is deferred to Appendix~\ref{app:bobwlowebound}).
\begin{theorem}\label{thm:bobwlowerbound} The robustness of any deterministic clock auction that is augmented with a prediction $\pred$ and satisfies $(1+\epsilon)$-consistency for some constant $\epsilon>0$ is $\Omega(\frac{1}{\epsilon}\log n)$.
\end{theorem}
\subsection{The Error-Tolerant Auction}\label{sec:bobwerr}

We now demonstrate that Mechanism~\ref{alg:follow-the-leader} can be extended to achieve a constant approximation not only when the prediction is optimal, but also when the predicted set deviates from the optimal set by a constant factor in terms of value. To this end, we define the \emph{prediction error} of a prediction $\pred$ in a given instance $I$ as the multiplicative difference between the value of the predicted set and the value of the actual optimal set:
\begin{align}\label{def:error}
    \eta(\pred, I) = \frac{v(\opt(I))}{v(\pred)}.
\end{align}
This type of error is called the \emph{quality of recommendation} by \cite{CSV24}.
Next, we introduce the \errmech\ auction, which is an extension of Mechanism~\ref{alg:follow-the-leader}. This auction takes as input a parameter $\tol \geq 1$, referred to as the error-tolerance parameter, which is chosen by the mechanism designer. The only modification from Mechanism~\ref{alg:follow-the-leader} to \errmech\ is that in Line~\ref{alg:line:Prevc}, the targeted revenue is adjusted from $R_t$ to $R_t/\tol$. When the prediction error $\eta$ is at most $\tol$, \errmech\ achieves an approximation of $(1+\epsilon)\eta$ to the optimal welfare. Additionally, it maintains an approximation of $\left(\frac{2000\tol(1+\epsilon)}{9\epsilon}+1\right)H_n$ to the optimal welfare at all times. Note that if $\tol = 1$, the auction reduces to Mechanism~\ref{alg:follow-the-leader}, preserving the corresponding consistency and robustness guarantees. The formal description of \errmech\ and the proof of the following theorem are deferred to Appendix~\ref{app:bobwerr}.

\begin{theorem}\label{thm:bobw-err}
    For any downward-closed set system and predicted optimal set, \errmech\ with parameter $\gamma = \frac{10(1+\epsilon)}{9\epsilon}$ and $\tol \geq 1$ achieves an approximation of $(1+\epsilon)\eta$ if $\eta \leq \tol$ and $\left(\frac{2000\tol(1+\epsilon)}{9\epsilon}+1\right)H_n$ otherwise, where $\eta$ is the prediction error defined in~\eqref{def:error}.
\end{theorem}

\section{Clock Auctions Guaranteeing a Stronger Notion of Consistency}\label{sec:strongerconsistency}

In this section, we turn to the more demanding notion of \sconsistency, which requires a good approximation of the predicted solution irrespective of the prediction quality. We show that achieving good \sconsistency\ is much harder than achieving good consistency. In particular, there is a separation in terms of the robustness that one can achieve for each of these two notions of consistency. Specifically, we show that if one wants \(\alpha\)-\sconsistency\ for some constant \(\alpha\), not only is it impossible to achieve \(O(\log n)\) robustness, as we achieved for the standard notion of consistency in the previous section, but one must actually suffer \(\Omega\left(n^{\frac{1}{\alpha-1}}\log n\right)\) robustness. Before proving this separation, we provide a learning-augmented clock auction that matches this bound for any constant $\alpha$.

\begin{theorem}\label{thm:strongconsistency}
    For any downward-closed set system, Mechanism \ref{alg:strong-consistency} with parameters $\consistency \in [1+\epsilon, H_n]$  and $$\robustness \geq \frac{ 8H_n \Gamma\left(n+\frac{\consistency}{\consistency - 1}\right)}{\Gamma\left(1+\frac{\consistency}{\consistency - 1}\right)n!}-\frac{4H_n(\alpha-1)}{\alpha} = \Theta\left(n^{\frac{1}{\consistency - 1}}\log n \right),$$ 
    where $\Gamma(x) = \int_0^{\infty}{t^{x-1}e^{-t}dt}$, is $\consistency$-consistent$^\infty$ and $\robustness$-robust.
\end{theorem}

\begin{figure}[h]
    \centering
    \includegraphics[width = 0.8 \textwidth]{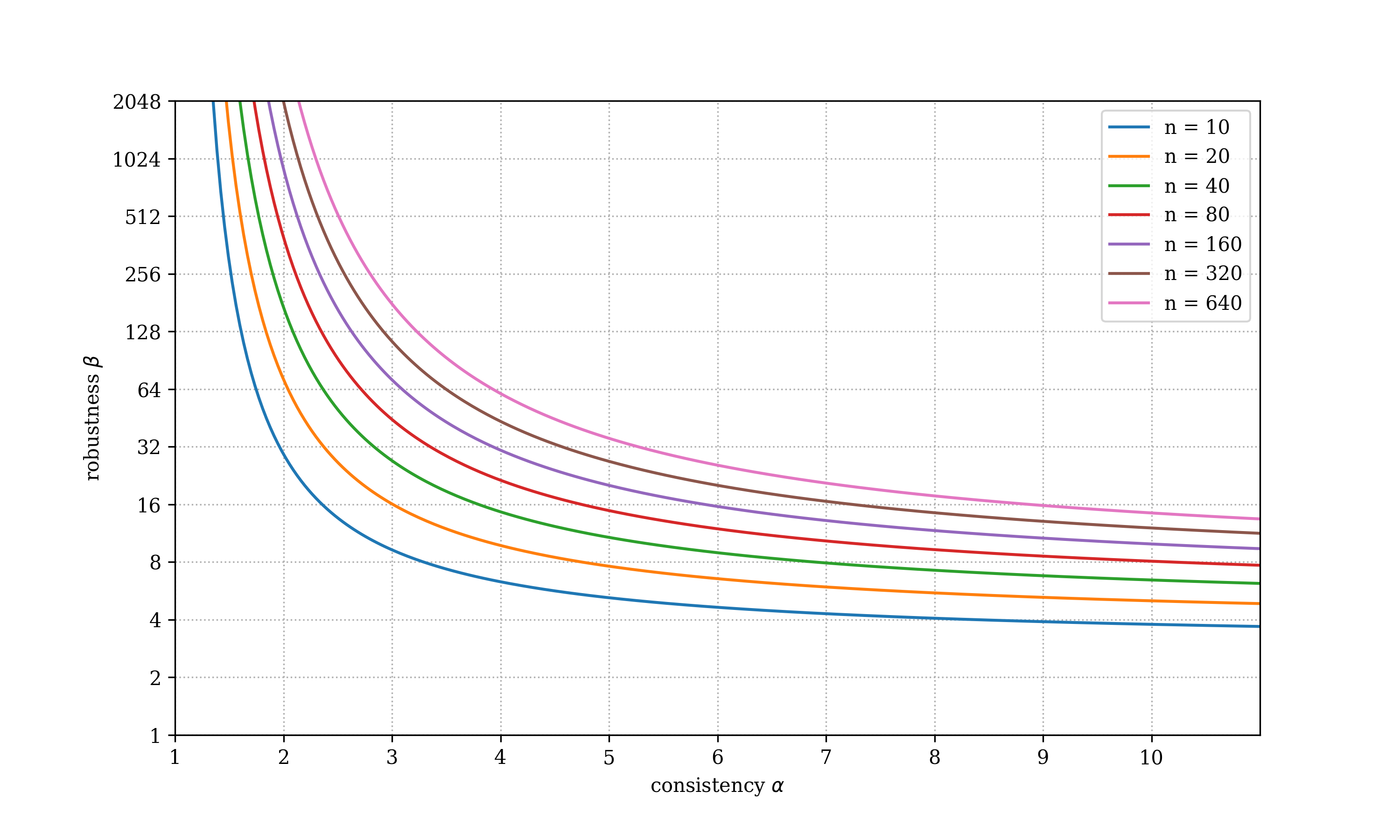}
        \caption{The trade-off between robustness and \sconsistency\ when $\beta=n^{1/(\consistency-1)}H_n$ for different values of $n$}
    \label{fig:sconsistencyplot}
\end{figure}

The clock auction defined in Mechanism \ref{alg:strong-consistency}, named \textsc{FollowTheBindingBenchmark}, also alternates between uniformly raising the bids of the unpredicted and predicted bidders while maintaining a careful balance between the revenue from the predicted set and the welfare of the unpredicted set. However, to achieve \sconsistency, it requires that in every round the revenue collected from the predicted set be at least \( \frac{1}{\alpha} \) of the rejected welfare of the unpredicted set plus the current revenue, i.e., the lower bound of the welfare of the predicted set. Unless this constraint is satisfied, the clock auction will continue raising the bids for the predicted set. See Figure~\ref{fig:sconsistencyplot} for a demonstration of the asymptotic trade-off the mechanism obtains for different sizes of instances.


\renewcommand*{\algorithmcfname}{MECHANISM}
\begin{algorithm2e}[h]
\DontPrintSemicolon
\LinesNumbered
\SetNoFillComment
\KwIn{A system $\mathcal{F}$ of feasible sets of bidders, a predicted optimal set $\pred$, a consistency parameter $\consistency$, a robustness parameter $\robustness$}

$F \gets F \setminus \{F \cap \pred\}$ for all $F \neq \pred \in \feasible$ \tcp*{Make $\pred$ disjoint from all unpredicted sets} \label{alg:line:make-disjoint-strong}
$p_i \gets v_{\text{min}}$ for all $i \in N$\tcp*{Raise price of each bidder to the initial small amount}

$\revP_0 \gets \rev(\pred, \price)$

$S \gets \pred$

$t \leftarrow 1$

\While{TRUE}{
Run $\uniform(N \setminus \pred, \price)$ until:(i) $\max_{F \in \feasible} \rev(F \cap N,\price) \geq \frac{\robustness}{4}\cdot \revP_{t-1}$; \textbf{OR} (ii) $\max_{F \in\feasible}v(F \setminus N) \geq \frac{\robustness}{4} \cdot \revP_{t-1}$ and $\max_{F \in \feasible} \rev(F \cap N,\price) \geq \frac{\robustness}{4H_n}\cdot \revP_{t-1}$ \label{alg:line:unpredictedphase}

\lIf{$N = \pred$}{
    \Return{$\pred$, $\price$}\label{alg:line:servePredicted}
    \tcp*{All unpredicted bidders reject case}}
$\tilde{R}^P_t \leftarrow \revP_{t-1} \cdot 2$ \label{alg:line:doubling}

Run $\uniform(\pred,\price)$ until: (i) $\rev(\pred, \price) \geq \tilde{R}^P_t$ \textbf{AND} (ii) $(\consistency-1) \cdot\rev(\pred, \price) \geq v(S \setminus \pred)$\label{alg:line:predictedphase} 

\lIf{$\pred = \emptyset$}{\Return{WFCA$(N,\price)$}\label{alg:line:WFCA2}
\tcp*{All predicted bidders reject case}}
$\revP_t \gets \rev(\pred,\price)$

$t \leftarrow t + 1$

}
\SetAlgoRefName{2}
\caption{\textsc{FollowTheBindingBenchmark}}
\label{alg:strong-consistency}
\end{algorithm2e}

As before, we assume that $\mathcal{F}$ is a set system where the predicted set is completely disjoint from all maximal unpredicted sets; by Observation~\ref{lem:make-disjointc}, this is without loss of generality. We first prove the robustness guarantee. All the missing proofs can be found in Appendix \ref{app:stronger}.

\subsubsection*{Robustness Argument.} To demonstrate that  Mechanism \ref{alg:strong-consistency} achieves the desired robustness we consider two cases depending on whether or not the auction terminates by serving a subset of the predicted set or a subset of some unpredicted set.  Since we assume that $\consistency < \robustness$ (and we will eventually show that our mechanism achieves social welfare within a $\consistency$-factor of the welfare in the predicted set), we show that our mechanism obtains welfare within a $\robustness$-factor of the best unpredicted set. Building toward this guarantee, we first prove a useful lemma regarding the total welfare lost from an unpredicted set.

\begin{lemma}\label{lem:unpred-welf-rejected}
    The total welfare rejected in any unpredicted set $S$ in some iteration $t$ is at most $\frac{\robustness}{2} \cdot \revP_{t-1}.$
\end{lemma}

As a corollary of this lemma, we can bound the total welfare lost from any unpredicted set of bidders by observing that the revenue and welfare targets increase by at least a factor $2$ in each round.

\begin{corollary}\label{cor:welf-rejected-unpredicted}
    The total welfare rejected in any unpredicted set $S$ throughout the first $t'$ iterations is at most $\robustness \revP_{t'-1}$.
\end{corollary}

 We are now ready to prove the robustness guarantees of Mechanism \ref{alg:strong-consistency}. Similarly as in Section~\ref{sec:bobw} we will analyze the two cases which depend on whether or not a subset of $\pred$ is output separately. We first consider the case where a subset of $\pred$ is output. The intuition is that we only serve a subset of predicted bidders if we do not satisfy either condition in Line \ref{alg:line:unpredictedphase} but we \emph{secured} good revenue from the predicted set in the previous iteration of the while loop.  We formalize this intuition in the following lemma.

\begin{lemma}
If Mechanism \ref{alg:strong-consistency} outputs a subset of the predicted set, then Mechanism \ref{alg:strong-consistency} is $\beta$-robust.
\end{lemma}
\begin{proof}
    Let $\hat{t}$ denote the value of $t$ when the auction terminates.  By assumption that we serve a subset of the predicted set and definition of the mechanism, this was a result of not satisfying either condition in Line \ref{alg:line:unpredictedphase}.  Note that we proceeded to round $\hat{t}$ because we satisfied the conditions of Line~\ref{alg:line:predictedphase} in round $\hat{t}-1$.  As such, the welfare remaining in the the predicted set is at least $\revP_{\hat{t}-1}$ (the revenue of the predicted set at the end of the previous round).  On the other hand, we know that the total welfare in any unpredicted set (since all unpredicted bidders were rejected) is at most $\robustness \revP_{\hat{t}-1}$ by Corollary \ref{cor:welf-rejected-unpredicted}.  Therefore we obtain welfare within a $\robustness$ fraction of the optimal unpredicted set.
\end{proof}

We now turn toward the case where a subset of an unpredicted set is output. As in Section \ref{sec:bobw}, we consider the welfare contributions from bidders who exit the auction before the WFCA and those active during the WFCA separately.  

\begin{lemma}
If Mechanism \ref{alg:strong-consistency} outputs a subset of some unpredicted set, then Mechanism \ref{alg:strong-consistency} is $\beta$-robust.
\end{lemma}
\begin{proof}
    Let $\hat{t}$ denote the value of $t$ when the auction terminates.  By assumption that we serve a subset of the unpredicted set and definition of the mechanism, the auction terminates in Line \ref{alg:line:WFCA2}.  We may divide the welfare of the best unpredicted set into two components -- the portion of the welfare coming from bidders rejected during before the WFCA is run and the portion of the welfare coming from bidders which are active when the WFCA is run in Line \ref{alg:line:WFCA2}.  By Corollary \ref{cor:welf-rejected-unpredicted} we have that the total welfare rejected in any unpredicted set through the first $\hat{t}$ iterations is $\robustness \revP_{\hat{t} - 1}$. 
    On the other hand, we have that the revenue of the best unpredicted set before we run Line \ref{alg:line:WFCA2} is at least $\frac{\robustness}{4H_n} \cdot \revP_{\hat{t} -1 }$.  In addition, by Lemma \ref{lem:WFCA-rev-monotone} 
    we know that the revenue after completing Line \ref{alg:line:WFCA2} must then be at least $\frac{\robustness}{4H_n} \cdot \revP_{\hat{t} -1 }$.  This gives that the revenue collected by Mechanism \ref{alg:strong-consistency} is within a $4H_n$-factor of the welfare in the best unpredicted set contributed from bidders rejected before Line \ref{alg:line:WFCA2} is run.  
    Moreover, we have by Lemma \ref{lem:WFCA-rev-monotone} that the welfare achieved by our auction is within a $2H_n$-factor of the welfare in the best unpredicted set contributed from bidders who are active when Line \ref{alg:line:WFCA2} is run.  Combining these we have that if a subset of the unpredicted bidders is served then Mechanism \ref{alg:strong-consistency} obtains welfare within a factor $4H_n + 2H_n \leq \robustness$ of the best unpredicted set.
\end{proof}

\subsubsection*{Consistency Argument.} We now turn toward the more technically challenging \sconsistency\ guarantee. Recall again that a mechanism achieves $\consistency$-\sconsistency\ if it achieves an $\consistency$-approximation to the value of the \emph{predicted} set, irrespective of the prediction quality. To this end, we first resolve the simpler case where we serve some subset of predicted set. In this case the guarantee is achieved by design, but we include a formal proof of this for completeness in Appendix \ref{app:stronger}.
\begin{lemma}\label{lem:strong-cons-pred}
    If Mechanism \ref{alg:strong-consistency} outputs a subset of the predicted set, then Mechanism \ref{alg:strong-consistency} is $\consistency$-consistent$^\infty$.
\end{lemma}

The more difficult case involves ensuring $\consistency$-\sconsistency\ given that we output a subset of some unpredicted set. Intuitively we want to argue that by defining a large enough $\beta$, the revenue we obtained from the best unpredicted set is enough to ``cover'' the maximum potential welfare the predicted set can contain. To this end, we begin with some individual bounds of bidders' values in the predicted set based on when they rejected the price offer.

\begin{lemma}\label{lem:obsmerge}
Consider any iteration $t$. At the start of running Line~\ref{alg:line:predictedphase}, condition (ii) is true. In addition, while (ii) remains true, the current price offered to the bidders is less than $\tilde{R}_{t}^P/k$ where $k$ is the number of active bidders in the predicted set.
\end{lemma}
\begin{proof}
We begin by demonstrating the first part of the lemma statement. First observe that, by design, we satisfied both conditions (i) and (ii) of Line \ref{alg:line:predictedphase} at the end of iteration $t-1$.  Furthermore, since we do not raise any prices for predicted bidders in iteration $t$ before reaching Line \ref{alg:line:predictedphase} again, it must be that in iteration $t$ we satisfy condition (ii) at the start of Line \ref{alg:line:predictedphase}.  For the second part of the lemma statement, suppose that condition (ii) of Line \ref{alg:line:predictedphase} is satisfied and $k$ bidders are active.  If we continue to uniformly raise prices for these $k$ bidders, then it must be we do not satisfy condition (i).  As such, it must be that these bidders have not yet accepted a price of $\tilde{R}_t^P/k$.
\end{proof}

\begin{claim}\label{cl:value-bounds}
    Fix a set of bidders $S$ and index  bidders in $S$ in non-increasing value order.  For all integers $1 \leq k < |S|$, the following two conditions hold: 
    \begin{enumerate}
        \item Suppose that
    $(\consistency -1)k v_k \leq \sum_{i = k+1}^{|S|}{v_i}$.  Then, if $(\consistency-1)(k-1)v_{k-1} \geq \sum_{i=k}^{|S|}{v_i}$ we have that \[v_{k-1} \geq \frac{(\consistency - 1)k + 1}{(\consistency - 1)(k-1)}v_k.\]
    \item Suppose that 
    $(\consistency -1)k v_k \geq \sum_{i = k+1}^{|S|}{v_i}$.  Then, if $(\consistency-1)(k-1)v_{k-1} \leq \sum_{i=k}^{|S|}{v_i}$ we have that \[v_{k-1} \leq \frac{(\consistency - 1)k + 1}{(\consistency - 1)(k-1)}v_k.\]
    \end{enumerate}
\end{claim}
\begin{proof}
    We show part 1 of the claim as part 2 follows, symmetrically, by flipping the inequalities. By assumption, we have that $\sum_{i = k+1}^{|S|}{v_i} \geq (\consistency -1)k v_k$ so substituting gives
    \begin{align*}
        (\consistency-1)(k-1)v_{k-1} &\geq \sum_{i=k}^{|S|}{v_i}
        \geq v_k + (\consistency - 1)kv_{k},
    \end{align*}
    or, equivalently,
    \[v_{k-1} \geq \frac{(\consistency - 1)k + 1}{(\consistency -1)(k-1)}v_k,\] as desired.
\end{proof}

\begin{lemma}\label{lem:value-upper-bound-consistency}
Consider any point when we are running \textsc{UniformPrice} on the predicted set (i.e., we are running Line~\ref{alg:line:predictedphase}) where condition (ii) is satisfied and let $k$ denote the number of active bidders in the predicted set.  If the smallest valued predicted bidder with value $p_k$ rejecting causes condition (ii) to be violated, then if condition (ii) is not satisfied before the $k-1$-th largest bidder exits the auction her value $p_{k-1}$ is upper bounded by
\begin{equation}
    p_{k-1} < \frac{(\consistency - 1)k + 1}{(\consistency - 1)(k-1)} \cdot p_k.
\end{equation}
\end{lemma}
\begin{proof}
    Fix some point in the auction when the revenue of the predicted set satisfied $\consistency$-\sconsistency, i.e., $(\consistency - 1)$ times the current predicted revenue is greater than or equal to the rejected welfare in the predicted set, and let $k$ denote the number of remaining active predicted bidders and $\overline{W}$ denote the total rejected welfare at this point. Let $p_k$ denote the price offered to these $k$ active bidders just before one of them exits the auction. We then have $(\consistency - 1) \cdot kp_k \geq \overline{W}$.  If the lowest valued bidder among these $k$ exiting the auction causes $\consistency$-\sconsistency\ to be violated, i.e., $\consistency \cdot (k-1)p_k < p_k + \overline{W}$ we may obtain an upper bound on the price $p_{k-1}$ corresponding to the point at which $\consistency$-\sconsistency\ would be  satisfied again if $k-1$ bidders accept $p_{k-1}$.
    By applying part 2 of Claim \ref{cl:value-bounds}, we obtain that  
    \begin{equation*}
    p_{k-1} < \frac{(\consistency - 1)k + 1}{(\consistency - 1)(k-1)} \cdot p_k,
    \end{equation*}
    as desired.
\end{proof}

    

\begin{observation}\label{obs:consistency-lb-more}
    For any $k>2$ and $\consistency > 1$ we have that \[\frac{1}{k-1} \leq \frac{1}{k}\cdot \frac{(\consistency - 1)k + 1}{(\consistency-1)(k-1)} = \frac{1}{k-1} + \frac{1}{k(\consistency - 1)(k-1)}.\]
\end{observation}

With Lemma \ref{lem:value-upper-bound-consistency} and Observation \ref{obs:consistency-lb-more} in hand we can now move to show that the auction achieves $\consistency$-\sconsistency\ when a subset of the unpredicted set is output provided we have $\robustness$ large enough.  A crucial function in both the upper and lower bounds we demonstrate is $\Gamma(x)$ where $\Gamma(x) = \int_0^{\infty}{t^{x-1}e^{-t}dt}$ is the extension of the factorial function to the complex numbers, i.e., $\Gamma(k) = (k-1)!$ for $k \in \mathbb{N}$.

\begin{lemma}\label{lem:robustness_bound}
If Mechanism \ref{alg:strong-consistency} with parameters $\consistency \in [1+\epsilon, H_n]$  and $$\robustness \geq \frac{ 8H_n \Gamma\left(n+\frac{\consistency}{\consistency - 1}\right)}{\Gamma\left(1+\frac{\consistency}{\consistency - 1}\right)n!}-\frac{4H_n(\alpha-1)}{\alpha} = \Theta\left( n^{\frac{1}{\consistency - 1}}\log n\right)$$ outputs a subset of some unpredicted set, then it is $\consistency$-consistent$^\infty$. 
\end{lemma}
\begin{proof}
     Consider the round $\hat{t}$ in which the auction terminates and assume that the auction proceeds to Line \ref{alg:line:WFCA2}, i.e., a subset of the unpredicted bidders is served.  By 
     Lemma~\ref{lem:obsmerge}, note that the revenue $\revP_{\hat{t} - 1}$ 
     reached by the predicted set in iteration $\hat{t} - 1$ was such that $(\consistency - 1) \cdot \revP_{\hat{t} - 1}$ was more than the total rejected welfare in the predicted set throughout the first $\hat{t} - 1$ iterations of the while loop, i.e., the iteration $\hat{t}-1$ concluded with the predicted set satisfying $\consistency$-consistency$^\infty$. 
     On the other hand, since we proceeded past Line \ref{alg:line:unpredictedphase} we have that the revenue in the unpredicted set reached at least $\frac{\robustness}{4H_n} \cdot \revP_{\hat{t} - 1}$.  We just then want to bound the total welfare lost from the predicted set in the final iteration $\hat{t}$.  
     By 
     Lemma~\ref{lem:obsmerge}
     we have that the first rejected bidder has value at most $2\revP_{\hat{t}-1}/k$ where $k$ is the number of active predicted bidders at the start of iteration $\hat{t}$.  Applying
     Lemma~\ref{lem:value-upper-bound-consistency} and Observation \ref{obs:consistency-lb-more} 
     to upper bound the value $v_{k-1}$ of the next rejecting bidder, we obtain that \[v_{k-1} \leq \frac{2\revP_{\hat{t} - 1}}{k} \cdot \frac{(\consistency - 1)k + 1}{(\consistency - 1)(k-1)}.\]
     Iteratively applying Lemma~\ref{lem:value-upper-bound-consistency} and Observation \ref{obs:consistency-lb-more} to find an upper bound to each rejected bidder's value, we have that the total welfare rejected in this round is at most
    {\allowdisplaybreaks{
    \begin{align*}
        \frac{2\revP_{\hat{t}-1}}{k} + \sum_{i = 2}^{k}{  \frac{2\revP_{\hat{t}-1}}{k} \cdot \prod_{j = i}^{k} \frac{(\consistency -1)j + 1}{(\consistency - 1)(j-1)}}  &\leq \frac{2\revP_{\hat{t}-1}}{n} + \sum_{i = 2}^{n}{  \frac{2\revP_{\hat{t}-1}}{n} \cdot \prod_{j = i}^{n} \frac{(\consistency -1)j + 1}{(\consistency - 1)(j-1)}} \\
        &=\frac{2\revP_{\hat{t}-1}}{n} + \frac{2\revP_{\hat{t}-1}}{n} \cdot \sum_{i = 2}^{n}{\prod_{j = i}^{n} \frac{j + 1/(\consistency -1)}{j-1}}\\
        &=\frac{2\revP_{\hat{t}-1}}{n} + \frac{2\revP_{\hat{t}-1}}{n} \cdot \sum_{i = 2}^{n}{\frac{\Gamma(i-1)\Gamma\left(n + 1 + \frac{1}{\consistency-1}\right)}{\Gamma\left(\frac{1}{\consistency - 1} + i\right)\Gamma(n)}}\\
        &= \frac{2\revP_{\hat{t}-1}}{n} + \frac{2\revP_{\hat{t}-1}}{n} \cdot \left(-\consistency n + \frac{\consistency \Gamma\left(\frac{n\consistency + \consistency - n}{\consistency - 1}\right)}{\Gamma\left(\frac{2\consistency -1}{\consistency - 1}\right)\Gamma(n)} + n - 1\right)\\
        &= \frac{2\revP_{\hat{t}-1}}{n} \cdot \left(-\consistency n + \frac{\consistency \Gamma\left(\frac{n\consistency + \consistency - n}{\consistency - 1}\right)}{\Gamma\left(\frac{2\consistency -1}{\consistency - 1}\right)(n-1)!} + n\right),
    \end{align*}}}
    where the inequality comes from the fact that for any $\consistency > 1$ the partial product from $j = k+1$ to $n$ times $2\revP_{\hat{t}-1}/n$ yields a value at least $2\revP_{\hat{t}-1}/k$, the second equality comes from the fact that $z\cdot \Gamma(z) = \Gamma(z+1)$ for all $z \in \mathbb{R}$ and, hence, $\Gamma(x+y)/\Gamma(x) = x \cdot (x+1) \cdot \dots \cdot (x+y-1)$ for all $y \in \mathbb{N}$ and all $x\in \mathbb{R}$, and the third equality is due to Lemma \ref{lem:summation-lemma} which we defer to Appendix \ref{sec:aux-lemmas}.
    
    Combining the welfare rejected from the first $\hat{t} - 1$ rounds and the welfare from the last round we obtain that the total welfare rejected from the predicted set is at most \[(\consistency - 1) \cdot \revP_{\hat{t} - 1} + \frac{2\revP_{\hat{t}-1}}{n} \cdot \left(-\consistency n + \frac{\consistency \Gamma\left(\frac{n\consistency + \consistency - n}{\consistency - 1}\right)}{\Gamma\left(\frac{2\consistency -1}{\consistency - 1}\right)(n-1)!} + n\right).\]

    But then, since our auction reached revenue at least $\frac{\robustness}{4H_n} \cdot \revP_{\hat{t} - 1}$ before running the WFCA and the WFCA is revenue monotone by Lemma \ref{lem:WFCA-rev-monotone} it suffices that \[\alpha \frac{\robustness}{4H_n} \geq (\consistency - 1)  + \frac{2}{n} \cdot \left(-\consistency n + \frac{\consistency \Gamma\left(\frac{n\consistency + \consistency - n}{\consistency - 1}\right)}{\Gamma\left(\frac{2\consistency -1}{\consistency - 1}\right)(n-1)!} + n\right)\] or, equivalently,
    \[\robustness \geq \frac{4H_n}{\consistency}\cdot\left((\consistency - 1)  + \frac{2}{n} \cdot \left(-\consistency n + \frac{\consistency \Gamma\left(\frac{n\consistency + \consistency - n}{\consistency - 1}\right)}{\Gamma\left(\frac{2\consistency -1}{\consistency - 1}\right)(n-1)!} + n\right)\right).\]

Therefore, for the proof to go through it suffices for $\beta$ to satisfy
\[\robustness ~\geq~ \frac{4H_n}{\consistency} (\consistency-1-2\consistency +2) + \frac{ 8H_n \Gamma\left(n+\frac{\consistency}{\consistency - 1}\right)}{\Gamma\left(1+\frac{\consistency}{\consistency - 1}\right)n!} ~=~ \frac{ 8H_n \Gamma\left(n+\frac{\consistency}{\consistency - 1}\right)}{\Gamma\left(1+\frac{\consistency}{\consistency - 1}\right)n!}-\frac{4H_n(\consistency-1)}{\consistency}.\]
As we care about the asymptotic behavior of this bound and $n + \frac{\consistency}{\consistency + 1}$ is a positive real number, we can apply Stirling's approximation, i.e., \[\Gamma(z) = \sqrt{\frac{2\pi}{z}}\left(\frac{z}{e}\right)^z\left(1+O\left(\frac{1}{z}\right)\right),\] to obtain that the proof goes through with
\begin{align*}
\robustness &= \frac{8H_n}{\Gamma\left(1 + \frac{\alpha}{\alpha - 1}\right)} \cdot \frac{\sqrt{\frac{2\pi}{n+1+1/(\consistency - 1)}} \left(\frac{n+1+1/(\consistency - 1)}{e}\right)^{n+1+1/(\consistency - 1)} \left(1 + O\left(\frac{1}{n+1+1/(\consistency - 1)}\right)\right)}{\sqrt{\frac{2\pi}{n+1}}\left(\frac{n+1}{e}\right)^{n+1} \left(1 + O\left(\frac{1}{n+1}\right)\right)}\\
&= \frac{8H_n}{\Gamma\left(1 + \frac{\alpha}{\alpha - 1}\right)} \cdot O(1) \cdot \left(\frac{n+1+1/(\consistency - 1)}{e}\right)^{n+1+1/(\consistency - 1)} \cdot \left(\frac{e}{n+1}\right)^{n+1}\\
&= \frac{8H_n}{\Gamma\left(1 + \frac{\alpha}{\alpha - 1}\right)} \cdot O(1) \cdot \frac{(n + 1 + 1/(\consistency - 1))^{n+1 + 1/(\consistency - 1)}}{(n+1)^{n+1}} \cdot \frac{1}{e^{1/(\consistency - 1)}}\\
&=\Theta\left(n^{\frac{1}{\consistency - 1}}\log n\right)
\end{align*}
where the last line is since $\Gamma(c)$ is a constant for any constant real number $c$ and $H_n \geq \consistency > 1+\epsilon$ for constant $\epsilon > 0$ so $1/(\consistency - 1) = O(1)$ and $\Gamma(1 + \consistency/(\consistency - 1)) = O(1)$.
\end{proof}
\subsection{Lower Bound}\label{sec:strongerconsistencylowerbound}


We now provide a lower bound for all deterministic clock auctions in terms of the trade-off between \sconsistency\ and robustness. 
This lower bound shows that the trade-off achieved by Mechanism~\ref{alg:strong-consistency} is near-optimal; in particular for any constant $\alpha$, the trade-off achieved by the mechanism is asymptotically optimal.  We note that the lower bound below given in Theorem \ref{thm:bobwlowerbound-consistency2} is interesting only for $\alpha$ with $\alpha^\alpha = o(n)$.

\begin{theorem}\label{thm:bobwlowerbound-consistency2} The robustness of any deterministic clock auction that is augmented with a prediction $\pred$ and satisfies $\consistency$-\sconsistency for $\consistency\in [1+\epsilon, H_n]$ is $\Omega(\frac{1}{\consistency}n^{1/(\consistency-1)}\log{n})$.
\end{theorem}

Note that for \( \alpha = 2 \), we have \( \beta = \Theta(n \log n) \). As the \sconsistency\ guarantee \( \alpha \) approaches 1, the robustness grows exponentially (e.g., for \( \alpha = 4/3 \), \( \beta = \Theta(n^3 \log n) \)). This is in sharp contrast with the standard consistency notion, for which we achieve \( (1 + \epsilon) \)-consistency with \( O(\log n) \)-robustness for any arbitrarily small \( \epsilon \).

\paragraph{Implications of the impossibility result on the feasible smoothness guarantees.} Before presenting the proof of this impossibility result, we would like to first point out its implications regarding the ``price of smoothness,'' i.e., the best achievable robustness if we require that the performance of the auction degrades smoothly as a function of the prediction error. Although a lot of prior work in the learning-augmented framework aims to design mechanisms with smoothness guarantees, to the best of our knowledge, (almost) none of this work provides explicit impossibility results regarding the price of smoothness. In contrast to this paucity of impossibility results, the robustness lower bounds that we prove for $\alpha$-$\text{consistent}^{\infty}$ auctions imply that any clock auction whose performance degrades linearly as a function of the error, $\eta$, needs to be $\Omega(\text{poly}(n))$-robust. 

This bound on the price of smoothness is implied by the following general observation (which holds beyond our setting, as long as the error is measured as in \eqref{def:error}, termed \emph{quality of recommendation} by \cite{CSV24}: any mechanism whose performance degrades linearly as a function of the error $\eta$, e.g., guaranteeing an approximation factor of $\alpha \eta$ if the prediction error is $\eta$, also satisfies $\alpha$-\sconsistency. To verify this, note that the value of a predicted set $\pred$ with error $\eta$ is $\text{OPT}/\eta$ (by definition). A mechanism with approximation factor $\alpha\eta$ guarantees welfare $\frac{\text{OPT}}{\alpha\eta} = \frac{\pred}{\alpha}$, which implies a \sconsistency\ of $\alpha$. This leads to the following Corollary regarding the price of smoothness.
\begin{corollary}
If a deterministic clock auction augmented with a prediction $\pred$ guarantees an approximation of $\alpha \eta$ for every $\alpha \in [1+\epsilon, H_n]$, where $\eta$ is the prediction error (as defined in \eqref{def:error}), then its robustness cannot be better than $\Omega(\frac{1}{\consistency}n^{1/(\consistency-1)}\log{n})$.
\end{corollary}

We now provide the proof of Theorem~\ref{thm:bobwlowerbound-consistency2}.
\begin{proof}(\emph{Proof of Theorem~\ref{thm:bobwlowerbound-consistency2}}) Let $\alpha$ be the targeted consistency any mechanism is aiming for.
Consider a feasibility constraint $\feasible$ defined by two disjoint maximal feasible sets $F_1$ and $F_2$, i.e., $N= F_1 \cup F_2$, and a set $F\subseteq N$ is feasible if and only if $F\subseteq F_1$ or $F\subseteq F_2$. Let the sizes of these sets be $|F_1|=k_1$ and $|F_2|=\alpha k_2$, and let $F_2$ be the set predicted to be optimal, i.e., $\pred = F_2$.

Using this feasibility constraint, we now define a family of instances $\mathcal{I}$ by defining the set of values that can be assigned to the bidders in $F_1$ and the bidders in $F_2$; the class of instances in $\mathcal{I}$ corresponds to all possible assignments of these values to the bidders in the corresponding sets. Let 
\[V_1=\left\{1, \frac{1}{2}, \frac{1}{3}, \dots, \frac{1}{k_1}\right\}\] 
be a set of $k_1$ values that can be matched to the $k_1$ bidders in $F_1$ 
and let 
\[V_2=\{v_1, v_2, \dots, v_{k_2} \}\] 
be a set of $k_2$ values that can be assigned to the $\alpha k_2$ bidders in $F_2$. In particular, each value $v_i$ for $i \in [1, k_2-1]$ is assigned to a single bidder, while $v_{k_2}$ is assigned to the remaining $(\alpha-1)k_2 +1$ bidders. The values in $V_2$ are in decreasing sequence such that the highest value among them is $v_1=1$ and, inductively, for every $i\in \{2, 3, \dots, k_2\}$:
\begin{align}\label{eq:inductivechange}
    v_i = \frac{(\consistency-1)(i-1)+\delta}{(\consistency-1)i+1}\cdot v_{i-1},
\end{align}
where $\delta>0$ is some arbitrarily small constant. Note that as $\delta\to 0$ and $\consistency\to \infty$, this sequence of values converges to the harmonic sequence $\left\{1, \frac{1}{2}, \frac{1}{3}, \dots, \frac{1}{k_2}\right\}$. However, for smaller values of $\consistency$ the values in $V_2$ drop much faster than that. Specifically, the lowest value in $V_2$ is 
\begin{equation}\label{eq:smallest_value}
v_{k_2} ~=~ \prod_{i=2}^{k_2}  \frac{(\consistency-1)(i-1)+\delta}{(\consistency-1)i+1}~=~\frac{\Gamma\left(1+\frac{\consistency}{\consistency-1}\right)(k_2-1)!}{\Gamma\left(k_2+\frac{\consistency}{\consistency-1}\right)}+\delta',
\end{equation}
where $\delta'\to 0$ as $\delta\to 0$. 


By Equation \eqref{eq:inductivechange}, we have that $(\alpha - 1) \cdot (k_2-1) \cdot v_{k_2-1} < ((\alpha - 1)k_2 + 1)v_{k_2}$. Note that the right-hand side is the sum of the values of the bidders in $F_2$ with value $v_{k_2}$.  Again by Equation \eqref{eq:inductivechange} and part 1 of Claim \ref{cl:value-bounds} we have,
for every $i\in \{1, 2, \dots, k_2-1\}$ it satisfies:
\begin{equation}\label{eq:consistency_violation}
    (\consistency-1)\cdot i \cdot v_i < ((\consistency -1)k_2 + 1) v_{k_2} + \sum_{j=i+1}^{k_2-1} v_j.
\end{equation}

To prove this theorem, we consider any learning-augmented clock auction $\mech$ and simulate it on an instance from $\mathcal{I}$, chosen adversarially to maximize the number of bidders that drop out. Note that, since $\mech$ is deterministic, the adversary can essentially determine who to assign each value to after observing the price trajectory that $\mech$ will follow. Specifically, whenever $\mech$ raises the price of an active bidder $i\in F_1$ to an amount that exceeds some value $v\in V_1$ that remains active (i.e., that was not already assigned to a bidder that has dropped out), then the adversary can assign the value $v$ to $i$ and, as a result, $i$ drops out. This implies that, if at any point during the execution of $\mech$ the maximum price offered to an active bidder in $F_1$ is $p_1$, then the value of every active bidder in $F_1$ is at least $p_1$ (otherwise the adversary could have assigned that value to the bidder facing price $p_1$, causing him to drop out). Using the same argument for $F_2$, we can conclude that if at any point during the execution of $\mech$ the maximum price offered to an active bidder in $F_2$ is $p_2$, then the values of all active bidders in $F_2$ are at least $p_2$. 

We now partition the set of all clock auctions into three cases, based on their outcome facing the adversarially chosen instance from $\mathcal{I}.$

\textbf{Case 1 (the auction $\mech$ accepts a subset of $F_1$):} For this case, we show that there exists an instance (not in $\mathcal{I}$) for which the auction fails to achieve $\consistency$-\sconsistency. To verify this, let $x$ be the number of bidders in $F_1$ that did not drop out and note that, given the harmonic structure of the values in $V_1$, the smallest value among them will be at most $1/x$. As we showed above, this means that the maximum price offered to any one of the winning bidders in $F_1$ was $p_1\leq 1/x$. Now, consider an alternative instance that is identical in terms of the values of all the bidders that dropped out, but has value $1/x$ for all of the bidders that won. Note that the outcome of the auction would be the same for this alternative instance, since the only difference is regarding values of bidders that are higher than the price offered to them, so the welfare achieved by $\mech$ in this alternative instance would be $x\cdot \frac{1}{x}=1$. However, the social welfare of $\pred=F_2$ is more than $\alpha$, which is implied by Inequality~\eqref{eq:consistency_violation} for $i=1$ since the left-hand side of $(\alpha - 1)v_1 = \alpha -1$ is less than the sum of all the remaining values in $F_2$, which violates the $\consistency$-\sconsistency\ constraint.


\textbf{Case 2 (the auction $\mech$ accepts a subset of $\pred$ of fewer than $k_2$ bidders):}
By the fact that we may assign values adversarially to ensure that the first $(\alpha - 1)k_2 + 1$ bidders whose clocks are raised above $v_{k_2}$ have value $v_{k_2}$, we may assume that all bidders with value $v_{k_2}$ exit the auction.  For this case, we once again show that there exists an instance (not in $\mathcal{I}$) for which the auction fails to achieve $\consistency$-\sconsistency.  To verify this, let $x\leq k_2-1$ be the number of bidders in $\pred$ that did not drop out and note that, by the definition of the values in $V_2$, the smallest value among them is $v_x$. As we showed above, this means that the maximum price offered to any one of the winning bidders in $\pred$ was $p_2\leq v_x$. We, once again, consider the alternative instance that is identical in terms of the values of all the bidders that dropped out, but has value $v_x$ for all the bidders in $\pred$ that won. The outcome of the auction would not be affected by this change, so the resulting welfare would be $x\cdot v_x$ but, using Inequality~\eqref{eq:consistency_violation} for $i=x$ for any $x \leq k_2 -1$, this violates the $\consistency$-\sconsistency\ constraint since the left-hand side of Inequality~\eqref{eq:consistency_violation} is $(\consistency-1)$ times the welfare the auction obtains and the right-hand side is the rejected welfare from the predicted set.

\textbf{Case 3 (the auction $\mech$ accepts a subset of $\pred$ of at least $k_2$ bidders):}
For this case, note that by the adversarial assignment of values, the auction maximizes its welfare if no price offered to a bidder in $\pred$ exceeded $v_{k_2}$, otherwise the adversarial assignment would ensure that the corresponding bidder would have dropped out (and $\mech$ then would not accept all bidders of value $v_{k_2}$). Now consider an alternative instance that is identical with respect to the values of the bidders in $F_1$, but the values of everyone in $F_2$ are equal to $v_{k_2}$. Note that the outcome of the auction would be the same in this alternative instance, since the only difference is regarding values of bidders that are higher than the price offered to them. Therefore, using Equation~\eqref{eq:smallest_value}, we conclude that the welfare achieved by the auction in this instance would be 
\[\alpha \cdot k_2\cdot v_{k_2} = \alpha \cdot k_2\cdot \frac{\Gamma\left(1+\frac{\consistency}{\consistency-1}\right)(k_2-1)!}{\Gamma\left(k_2+\frac{\consistency}{\consistency-1}\right)}+\delta' = \frac{\alpha \Gamma\left(1+\frac{\consistency}{\consistency-1}\right) k_2!}{\Gamma\left(k_2+\frac{\consistency}{\consistency-1}\right)}+\delta',\]
whereas the welfare of $F_1$ is equal to $H_{k_1}=1+\frac{1}{2}+\dots + \frac{1}{k_1}$, leading to a robustness of
\[\beta = \Omega \left( \frac{ H_{k_1} \Gamma\left(k_2+\frac{\consistency}{\consistency - 1}\right)}{\alpha \Gamma\left(1+\frac{\consistency}{\consistency - 1}\right)k_2!} \right)= \Omega\left(\frac{1}{\alpha} k_2^{1/(\consistency-1)}\log{k_1} \right),\]
where the last equation is inferred by following the same steps as we did at the end of the proof of Lemma~\ref{lem:robustness_bound}. 
If we let $k_1 = n^c$ and $k_2 = \frac{n-n^c}{\alpha}$, this yields the claimed robustness lower bound.
\end{proof}

\bibliographystyle{abbrvnat}
\bibliography{biblio}

\appendix
\section{Missing Proofs from Section~\ref{sec:bobw}}\label{app:bobw}
\subsection{Proof of Lemma \ref{lem:welf-single-round-U2c}}
\begin{proof}
    Fix some set $S$ and consider the process of raising the price offered to bidders in this set as in line \ref{alg:line:Urevc}.  If at least $k$ bidders in $S$ accept a price of $R_t/k$ for any $k$ then we terminate this process (since the total revenue in $S$ would reach the  revenue target of $R_t$).  This would happen if the $k$-th highest value bidder in $S$ had value more than $R_t/k$.  As such, rejecting the $k'$-th highest value bidder in iteration $t$ of the while loop implies that this bidder has value at most $R_t/k'$.  On the other hand, since there are at most $n$ bidders contained in set $S$ the total value of the bidders rejected in iteration $i$ of the while loop is at most \[\sum_{k' = 1}^{n}{\frac{R_t}{k'}} = R_t \cdot H_n, \] completing the proof.
\end{proof}

\subsection{Proof of Lemma \ref{lem:total-rejected-Predc}}
\begin{proof}
    Following the same argument of Lemma~\ref{lem:welf-single-round-U2c} we first observe that the total welfare rejected from the predicted set in line \ref{alg:line:Prev} in any iteration $t'$ of the while loop of \cmech\ is at most $R_{t'}\cdot H_n$.  Taking a sum over all rounds from $1$ to $t$ we then obtain that the total welfare rejected is at most 
    \[\sum_{t' = 1}^{t}{R_{i'} \cdot H_n} = H_n \sum_{t' = 1}^{t}{10^{t'}} = H_n \cdot \frac{10^{t+1} - 1}{9} \leq R_t \cdot \frac{10H_n}{9}, \]
    as desired.
\end{proof}

\subsection{Proof of Corollary \ref{cor:approx-ftul}}
\begin{proof}
    Suppose we have an algorithm which gives a $\rho$-approximation to the underlying algorithmic problem of maximizing welfare/revenue on a fixed set of values.  We first argue that the consistency of \cmech\ when using this approximation algorithm (in place of an exact algorithm) remains $(1+\epsilon)$.  To see this, observe that we can exactly compute the revenue within the predicted set in every round in polynomial time since we can simply sum the prices of the bidders in the predicted set.  As such, the arguments in Lemma \ref{lem:ftl-consistentc} still hold when we use an approximation algorithm in Lines 7 and 9 of \cmech\ on the \emph{unpredicted} sets.  
    
    It remains to verify that the robustness of \cmech\ is $O(\rho\cdot\frac{1}{\epsilon}\log{n})$.  First observe that, by definition of the fact that we are using a $\rho$-approximate algorithm in Line 7, the maximum welfare rejected from any unpredicted set in iteration $t$ of the while loop of Mechanism \ref{alg:follow-the-leader} at the end of line 7 is $\rho\cdot 2R_t\gamma H_n$.  Similarly, by the same reasoning as in the proof of Lemma \ref{lem:welf-single-round-U2c}, the total welfare rejected from any unpredicted set in line 9 of Mechanism \ref{alg:follow-the-leader} in iteration $t$ is at most $\rho R_t \cdot  H_n$ (since $k$ bidders in any set accepting a price $\rho R_t/k$ would mean that the best revenue is weakly higher than $\rho R_t$ and a $\rho$-approximate algorithm would find a set with revenue at least $R_t$).  Combining these two gives that the total welfare rejected from any unpredicted set in any iteration $t$ is at most $R_t \cdot (2\gamma + 1)\rho H_n$.  As in Lemma \ref{lem:Robust-Pred-winsc}, summing over all rounds then gives that the total welfare rejected from any unpredicted set from rounds $1$ to $t$ is at most $R_t \cdot \frac{(2\gamma + 1) 10 \rho H_n}{9}$.

    Following similar reasoning to Lemmas \ref{lem:Robust-Pred-winsc} and \ref{lem:unpred-wins-robustc}, we consider cases depending on whether a predicted or unpredicted set wins. Let $\hat{t}$ denote the final iteration of the while loop of Mechanism \ref{alg:follow-the-leader}.  If Mechanism \ref{alg:follow-the-leader} outputs a subset of the predicted set, observe that the revenue of the predicted bidders is at least $R_{\hat{t}-1} = \frac{R_{\hat{t}}}{10}$.  But then, since the total welfare in any unpredicted set is at most $R_{\hat{t}} \cdot \frac{(2\gamma + 1)10 \rho H_n}{9}$ (since we rejected all the bidders in some round between $1$ and $\hat{t}$), we obtain that Mechanism \ref{alg:follow-the-leader} outputs a set with at least a $\frac{9}{\rho \cdot (2\gamma + 1)100H_n}$-fraction of the optimal welfare.  Suppose, instead, that Mechanism \ref{alg:follow-the-leader} outputs a subset of some unpredicted set (i.e., we reach line \ref{alg:line:WFCAc}) in the final iteration of the while loop $\hat{t}$.  By the same arguments in Lemma \ref{lem:welfare-before-WFCAc}, we know that the revenue among active bidders in the highest revenue set is $R_{\hat{t}}$ when we reach line \ref{alg:line:WFCAc}.  As above, the total welfare rejected from any unpredicted set by this point is at most $R_{\hat{t}} \cdot \frac{(2\gamma + 1)10 \rho H_n}{9}$.  But then, we have that the \emph{revenue} among active bidders in the highest revenue set when we reach line \ref{alg:line:WFCAc} is a $\frac{9}{\rho \cdot 10(2\gamma + 1)H_n}$-fraction of the maximum social welfare from any feasible set of rejected unpredicted bidders.  We can then apply similar reasoning to the proof of Lemma \ref{lem:unpred-wins-robustc} to complete the proof.  As noted in \cite{CGS22}, the WFCA can also readily admit the use of a $\rho$-approximation for maximizing the revenue at a fixed set of clock prices.  In this case, the approximation guarantee given the WFCA is $\rho \cdot 2H_n$ \cite[Theorem 5]{CGS22}.  Moreover, the WFCA remains revenue monotone even when using an approximation algorithm.  Combining these observations with our above reasoning, we have that the revenue from the set output by Mechanism \ref{alg:follow-the-leader} is a $\frac{9}{\rho \cdot 10(2\gamma + 1)H_n}$-fraction of the welfare rejected before running the WFCA and the welfare from the set output by Mechanism \ref{alg:follow-the-leader} is a $\frac{1}{\rho \cdot 2H_n}$.  Putting this all together, we have that Mechanism \ref{alg:follow-the-leader} with any $\gamma > 10/9$ and a $\rho$-approximate black-box achieves at least a $\frac{9}{\rho \cdot 10(2\gamma + 1)H_n} + \frac{1}{\rho \cdot 2H_n}$-fraction of the optimal welfare when the prediction is incorrect and Mechanism \ref{alg:follow-the-leader} outputs a subset of some unpredicted set.  Combining our two cases gives that Mechanism \ref{alg:follow-the-leader} is $O(\rho \cdot \frac{1}{\epsilon} \log{n})$-robust, as desired.
\end{proof}

\section{Missing Proof from Section~\ref{sec:bobwlowerbound}}\label{app:bobwlowebound}
\begin{proof}[Proof of Theorem~\ref{thm:bobwlowerbound}]
Consider any deterministic clock auction $\mech$ that is augmented with a prediction $\pred$ and achieves $(1+\epsilon)$-consistency and robustness better than $\Omega(\frac{1}{\epsilon}\log{n})$.
Also, consider a family of instances involving  an arbitrarily large 
number of bidders $n$ and a feasibility constraint $\feasible$ defined by two disjoint maximal feasible sets $F_1$ and $F_2$, i.e., $N= F_1 \cup F_2$, and a set $F\subseteq N$ is feasible if and only if $F\subseteq F_1$ or $F\subseteq F_2$. $F_1$ contains only a single bidder and $|F_2|=n-1$. Also, let $F_2$ be the set predicted to be optimal, i.e., $\pred = F_2$.

Using this feasibility constraint, we now define a family of instances $\mathcal{I}$ where the value of the single bidder in $F_1$ is $0.99$ and the values of the bidders in $F_2$ correspond to the following set of values:
\[V_2 = \left\{0.99, \frac{\epsilon}{2(H_{n-1}-1)}, \frac{\epsilon}{3(H_{n-1}-1)}, \dots, \frac{\epsilon}{(n-1)(H_{n-1}-1)}\right\}.\]

We now consider what $\mech$ does facing an instance in $\mathcal{I}$ that is chosen by adversarially matching the values in $V_2$ to bidders in $F_2$, depending on the deterministic price trajectory that $\mech$ will follow. Specifically, at each time $t$ during the execution of $\mech$, let $V_2(t)\subseteq V_2$ be the set of values from $V_2$ assigned to bidders that have not dropped out yet. Then, the adversarial assignment of values to bidders in $F_2$ ensure that if at any time $t$ the price offered to an active bidder $i\in F_2$ exceeds a value $v\in V_2(t)$, we can assume that the adversary assigns $v$ to $i$, so $i$ would drop out. As a result, if number of active bidders at any time $t$ is $x\geq 2$, then the highest value offered to any one of them is at most $\frac{\epsilon}{x (H_{n-1}-1)}$, so the revenue never exceeds $x\cdot \frac{\epsilon}{x (H_{n-1}-1)} =\frac{\epsilon}{H_{n-1}-1}$. 

Using this fact, with the assumption that the auction achieves better than $\Omega(\frac{1}{\epsilon}\log n)$, we can infer that the auction cannot terminate with at least two active bidder in $F_2$. Specifically, if $\mech$ terminated with $x\geq 2$ active bidders in $F_2$ and, as we argued above, their final price was at most $\frac{\epsilon}{x(H_{n-1}-1)}$, then $\mech$ would fail to satisfy the claimed robustness guarantee on an alternative instance (not in $\mathcal{I}$). This alternative instance is identical in terms of the values of all the bidders that dropped out and the only difference is in the value of the $x$ winning bidders, whose values are changed to $\frac{\epsilon}{x (H_{n-1}-1)}$. Note that the outcome of $\mech$ facing this alternative instance would be the same, since the prices offered to the winning bidders would still be below their values. However, in this alternative instance the welfare of $\mech$ would be $\frac{\epsilon}{H_{n-1}-1}$ while the optimal social welfare is $0.99$,  leading to a robustness of $\Omega(\frac{1}{\epsilon}\log{n})$.

If, on the other hand, the auction accepted at most 1 bidder from $F_2$, then the resulting welfare would be $0.99$. Consider the total value rejected so far, we would have:
$\sum^{n-1}_{i = 2} \frac{\epsilon}{ i \cdot (H_{n-1}-1)} = \epsilon,$
 since $\sum^{n-1}_{i=2} = H_{n-1}-1$. Therefore the consistency would be
 \[\frac{0.99+\epsilon}{0.99} > 1+\epsilon,\] violating the consistency guarantee.
\end{proof}
\section{Missing Proof from Section~\ref{sec:bobwerr}}\label{app:bobwerr}
For completeness we include the full description of \errmech\ below:
\begin{algorithm2e}[h]
\DontPrintSemicolon
\LinesNumbered
\SetNoFillComment
\KwIn{A system $\feasible$ of feasible sets of bidders, a predicted optimal set $\pred$, a parameter $\gamma$}

$F \gets F \setminus \{F \cap \pred\}$ for all $F \neq \pred \in \feasible$ \tcp*{Make $\pred$ disjoint from all unpredicted sets}
\label{alg:line:make-disjoint}

$p_i \gets v_{\text{min}}$ for all $i \in N$ \tcp*{Raise price of each bidder to the initial small amount}
$R_0 \gets \rev(\pred, \mathbf{p})$

$t \leftarrow 1$ \tcp*{Initialize the counter}

\While{TRUE}{
$R_t \gets 10^tR_{t-1}$

Run $\uniform(N \setminus \pred, \price)$ until $\max_Fv(F \setminus N) \in [R_t\gamma H_n,2R_t\gamma H_n)$\footnotemark
\tcp*{Reject ``safe'' amount of welfare to determine minimum total value in optimal set}\label{alg:line:Uwel}

\lIf{$N = \pred$}{
    \Return{$\pred$, $\price$ \label{alg:line:UwelTerm}}
}

Run $\uniform(N \setminus \pred, \price)$ until $\max_F\rev(F\cap N, \price) \geq R_t$ \tcp*{Ensure unpredicted sets ``cover'' their own lost welfare}\label{alg:line:Urev}

\lIf{$N = \pred$}{
    \Return{$\pred$, $\price$ \label{alg:line:UrevTerm}}
}
Run $\uniform(\pred, \price)$ until $\rev(\pred, \price) \geq R_t/\tol$ \label{alg:line:Prev} \tcp*{Ensure predicted set ``covers'' welfare lost from unpredicted sets}

\If{$\pred = \emptyset$}{
    \Return{ WFCA$(N, \price)$}\label{alg:line:WFCA}
}

$t \leftarrow t + 1$
}
\caption{\errmech}
\label{alg:errmech}
\end{algorithm2e}

\begin{lemma}\label{lem:ftl-consistent}
    Suppose that the predicted set has total welfare at least $1/\tol$ times the optimal social welfare (i.e., the prediction is within the error tolerance).  Then for a fixed parameter $\gamma > 10/9$ \errmech\ terminates in either  Line \ref{alg:line:UwelTerm} or Line \ref{alg:line:UrevTerm} and obtains at least a $\left(1 - \frac{10}{9\gamma}\right)$-fraction of the social welfare of the predicted set.
\end{lemma}
\begin{proof}
    Fix an instance $I$ where the optimal set has social welfare $W^*$ and where the predicted set has welfare at least $W^*/\tol$.  Consider the minimum value $k$ such that $10^{k} \cdot \gamma H_n$ exceeds $W^*$.  Let $i'$ denote the value of $i$ at the beginning of the final iteration of the while loop of \errmech\ when run on instance $I$.  Since we only continue to line \ref{alg:line:Urev} after the preceding if statement when there are still active unpredicted bidders it must be that $i' \leq k$ (as, otherwise, some unpredicted set would have total welfare greater than $W^*$) and if $i' = k$ then we terminate before line \ref{alg:line:Urev} (i.e., before we cause any additional predicted bidders to exit the auction in the iteration $k$).  
    
    Now consider the total amount of welfare lost from the predicted set from rounds $1$ to $i'-1$.  By Lemma \ref{lem:total-rejected-Predc} we have that this total welfare is at most $\frac{R_{i'-1}}{\tol} \cdot 10H_n/9 = \frac{10^{i' - 1}}{\tol} \cdot \frac{10H_n}{9} \leq \frac{10^{i^{*}-1}}{\tol} \cdot \frac{10H_n}{9}$.  On the other hand, by our definition of $k$, we know that the total social welfare in the predicted set is at least $\frac{10^{k-1}}{\tol} \cdot \gamma H_n$.  But then, at the end of the $i'-1$ iteration there must be active predicted bidders and moreover, we retain at least a 
    \[\frac{\frac{10^{k-1}}{\tol}\cdot \gamma H_n - \frac{10^{k-1}}{\tol}\cdot \frac{10}{9}H_n}{\frac{10^{k-1}}{\tol}\cdot \gamma H_n} = \frac{\gamma - \frac{10}{9}}{\gamma}\] fraction of the social welfare of the predicted set.
\end{proof}

\begin{lemma}\label{lem:Robust-Pred-wins}
    Consider some fixed constant parameter $\gamma > 10/9$ \errmech\ outputs a subset of the predicted set.  Then \errmech\ obtains social welfare within a $\frac{9}{\tol \cdot 100(2\gamma + 1)H_n}$-factor of the optimal social welfare.
\end{lemma}
\begin{proof}
    Consider the iteration $\hat{t}$ in which the while loop of \errmech\ terminates.  By definition, the mechanism terminates in either line \ref{alg:line:UwelTerm} or line \ref{alg:line:UrevTerm}.  By Line~\ref{alg:line:Uwel} and Lemma \ref{lem:welf-single-round-U2c} we have that the total welfare rejected from any unpredicted set in any round $\hat{t}$ is at most $2W_{\hat{t}} + R_{\hat{t}}H_n = R_{\hat{t}} \cdot (2\gamma + 1)H_n$.  But then, summing over all rounds we have that the total welfare contained in any unpredicted set is at most \[\sum_{\hat{t} = 1}^{\hat{t}}{R_{\hat{t}} \cdot (2\gamma + 1)H_n} = (2\gamma + 1)H_n \cdot \sum_{\hat{t} = 1}^{\hat{t}}{10^{\hat{t}}} = (2\gamma + 1)H_n \cdot \frac{10^{\hat{t} + 1} - 1}{9} \leq R_{\hat{t}} \cdot \frac{(2\gamma + 1)10H_n}{9}.\]  On the other hand, since we continued to iteration $\hat{t}$ we have that the revenue of the bidders in the predicted set reached $\frac{R_{\hat{t} - 1}}{\tol} = \frac{R_{\hat{t}}}{10\tol}$.  Moreover, this means that the welfare of the active predicted bidders is at least $\frac{R_{\hat{t}}}{10\tol}$ and we serve all these bidders.  Hence, \errmech\ obtains at least a $\frac{R_{\hat{t}}/(10\tol)}{R_{\hat{t}} \cdot \frac{(2\gamma + 1)10H_n}{9}} = \frac{9}{\tol \cdot 100(2\gamma + 1)H_n}$-fraction of the optimal social welfare.
\end{proof}


\begin{lemma}\label{lem:welfare-before-WFCA}
    Consider some parameter $\gamma > 10/9$ \errmech\ reaches line \ref{alg:line:WFCA}.  Then the \emph{revenue} of active bidders in the highest revenue set before running the Water-filling Clock Auction is at least a $\frac{9}{10(2\gamma + 1) H_n}$-approximation of the maximum social welfare obtainable by rejected unpredicted bidders.
\end{lemma}
\begin{proof}
    As before we consider the iteration $\hat{t}$ in which the while loop of \errmech\ terminates.  By assumption, our while loop terminates in line \ref{alg:line:WFCA}.  Following the same reasoning as Lemma \ref{lem:Robust-Pred-wins}, the total welfare rejected from any unpredicted set over all rounds from $1$ to $\hat{t}$ is at most \[\sum_{i' = 1}^{\hat{t}}{R_{i'} \cdot (2\gamma + 1)H_n} \leq R_{\hat{t}} \cdot \frac{(2\gamma + 1)10H_n}{9}.\]  On the other hand, because we arrive at line \ref{alg:line:WFCA} we know that there exists some set unpredicted set which obtains revenue $R_{\hat{t}}$ in round $\hat{t}$.  Thus, the revenue among active bidders in the highest revenue set when we reach line \ref{alg:line:WFCA} is a $\frac{9}{10(2\gamma + 1) H_n}$-fraction of the maximum social welfare obtained from any feasible set of rejected unpredicted bidders.
\end{proof}

\begin{lemma}\label{lem:unpred-wins-robust}
    Consider some parameter $\gamma > 10/9$ \errmech\ outputs a subset of unpredicted bidders.  Then \errmech\ obtains social welfare within a $\frac{9}{10(2\gamma + 1) H_n} + \frac{1}{4\log{n}}$-approximation to the optimal social welfare.
\end{lemma}
\begin{proof}
    In the case that \errmech\ outputs a subset of unpredicted bidders it must arrive at line \ref{alg:line:WFCA}.  As such, we can decompose the social welfare from the optimal (unpredicted) set into two components, the portion of the welfare coming from bidders rejected before \errmech\ runs the Water-filling Clock Auction and the portion of the welfare coming from bidders who participate in the WFCA.  From Lemma \ref{lem:welfare-before-WFCA} we know that the \emph{revenue} reached by \errmech\ before running the WFCA is within a $\frac{9}{10(2\gamma + 1) H_n}$-factor of the welfare rejected from the optimal set before running the WFCA.  Moreover, from Lemma \ref{lem:WFCA-rev-monotone} we have that the revenue reached by \errmech\ after running the WFCA is weakly higher than the revenue reached before running the WFCA.  As such, since the welfare obtained by serving a set of bidders is always weakly higher than the revenue collected from these bidders we have that the social welfare obtained by \errmech\ is within a $\frac{9}{10(2\gamma + 1) H_n}$-factor of the welfare rejected from the optimal set before running the WFCA.  Finally, we have from Lemma \ref{lem:WFCA-rev-monotone} that the social welfare achieved by running the WFCA only on bidders who are active when line \ref{alg:line:WFCA} is reached is within a $1/(4\log{n})$-factor of the optimal social welfare achievable from  these bidders.  Combining these two guarantees, we have that the social welfare obtained by \errmech\ is within a $\frac{9}{10(2\gamma + 1) H_n} + \frac{1}{4\log{n}}$-factor of the optimal social welfare.
\end{proof}
\begin{proof}[Proof of Theorem~\ref{thm:bobw-err}]
    First observe that $\epsilon > 0$ ensures that $\gamma > 10/9$.  In the case that the predicted set, indeed, has social welfare at least $1/\tol$ times the optimal social welfare then Lemma \ref{lem:ftl-consistent} gives that some subset of the predicted set is served.  Moreover, we obtain that the total social welfare of served bidders is within a factor \[\left(1 - \frac{10}{9\gamma}\right) = \left(1 - \frac{10}{9\frac{10(1+\epsilon)}{9\epsilon}}\right) = \frac{1}{1+\epsilon}\] of the total social welfare in the predicted set, thereby guaranteeing $(1+\epsilon)\eta$ approximation  with error tolerance $\tol$.  On the other hand, if the predicted set has welfare less than $1/\tol$ times the optimal (i.e., $\eta \leq \tol$) then we can consider two cases depending on whether or not a subset of the predicted set is ultimately served.  If some subset of the predicted set is served, Lemma \ref{lem:Robust-Pred-wins} guarantees that the total social welfare obtained by \errmech\ is within a factor $\frac{9}{\tol \cdot 100(2\gamma + 1)H_n}$ of the optimal social welfare.  Since $H_n = \Theta(\log{n})$ and $\gamma$ is a constant for any choice of constant $\epsilon$ we obtain $\Theta(\log{n})$-robustness in this case when $\tol = \Theta(1)$.  Finally, if some subset of an unpredicted set is served, Lemma \ref{lem:unpred-wins-robust} guarantees that \errmech\ obtains welfare within a factor $\frac{9}{10(2\gamma + 1)H_n} + \frac{1}{4\log{n}}$ of the optimal social welfare.  Again, since $H_n = \Theta(\log{n})$ and $\gamma$ is constant for any constant $\epsilon$ we obtain $\Theta(\log{n})$-robustness in this case, completing the proof.
\end{proof}

\section{Missing Proofs from Section~\ref{sec:strongerconsistency}}\label{app:stronger}
\subsection{Proof of Lemma \ref{lem:unpred-welf-rejected}}
\begin{proof}
    Fix a round $t$ in our auction and consider the total welfare lost by some unpredicted set $S$ in Line \ref{alg:line:unpredictedphase} during $t$.  First observe that if we move to Line \ref{alg:line:predictedphase} as a result of satisfying the second condition of the price increase for the unpredicted sets then it must be that every unpredicted set has rejected welfare less than $\frac{\robustness}{4} \cdot \revP_{t-1}$ (since we would have otherwise have stopped raising prices sooner, having met the lower revenue target of $\frac{\robustness}{4H_n}\cdot \revP_{t-1}$ before reaching $\frac{\robustness}{2}\cdot \revP_{t-1}$).  On the other hand, suppose we move to Line \ref{alg:line:predictedphase} as a result of satisfying the first condition.  Once some unpredicted set has rejected welfare $\frac{\robustness}{4}\cdot \revP_{t-1}$, this condition will be satisfied if there exists any set with $k$ remaining bidders of value at least $\frac{\robustness}{4H_n}\cdot \frac{\revP_{t-1}}{k}$.  As such, if we reject the $k$-th highest value bidder from any unpredicted set in round $t$ after having rejected after having rejected $\frac{\robustness}{4}\cdot \revP_{t-1}$ from some unpredicted set it must be that her value was at most $\frac{\robustness}{4H_n}\cdot \frac{\revP_{t-1}}{k}$. However, since each set has size at most $n$ the total value rejected from any unpredicted set $S$ in round $t$ after having rejected welfare at least $\frac{\robustness}{4}\cdot \revP_{t-1}$ from some unpredicted set is at most \[\sum_{k = 1}^{n}{\frac{\robustness}{4H_n}\cdot \frac{\revP_{t-1}}{k}} = \frac{\robustness}{4} \cdot \revP_{t-1}.\]  In total, we cannot reject total value from any set more than $\frac{\robustness}{2}\cdot \revP_{t-1}$. Similarly, if neither condition is met (i.e., we reject all unpredicted bidders before reaching either (i) or (ii) in Line \ref{alg:line:unpredictedphase}) we cannot reject  total value from any set more than $\frac{\robustness}{4}\cdot \revP_{t-1} + \frac{\robustness}{4}\cdot \revP_{t-1} = \frac{\robustness}{2}\cdot \revP_{t-1}$.
\end{proof}

\subsection{Proof of Corollary \ref{cor:welf-rejected-unpredicted}}
\begin{proof}
    Consider that in each round $\tilde{R}_t^P$ is set to double of $\revP_{t-1}$ in Line \ref{alg:line:doubling}.  And, by the conditions of Line \ref{alg:line:predictedphase} we have that the value of $\revP_t$ at the end of iteration $t$ in the while loop is at least twice $\revP_{t-1}$.  Applying Lemma \ref{lem:unpred-welf-rejected} to each round $t < t'$ we then have that the total welfare rejected through the first $t'$ iterations from any unpredicted set is at most \[\sum_{t = 1}^{t'}{\frac{\robustness}{2}R^P_{t-1}} = \sum_{t = 1}^{t'-1}{\frac{\robustness}{2}R_{t-1}^P} + \frac{\beta}{2}R^P_{t'-1} \leq \robustness R_{t'-1}^P,\] as desired.
\end{proof}

\subsection{Proof of Lemma \ref{lem:strong-cons-pred}}
\begin{proof}
    Observe that when we terminate, by assumption that we output a subset of the predicted set, we did so because we did not satisfy either condition in Line \ref{alg:line:unpredictedphase}.  But in the previous iteration of the while loop we have that the revenue of the predicted set was $(\consistency - 1)$ times the total welfare rejected from the predicted set.  The social welfare remaining in the predicted set is then at least a $\consistency$-factor of the total welfare of the predicted set.
\end{proof}

\section{Auxiliary Lemmas}\label{sec:aux-lemmas}
\allowdisplaybreaks
\begin{lemma}\label{lem:summation-lemma}
    For any choice of $\consistency > 1$ and all $n>2$, we have that \[\sum_{i = 2}^{n}{\frac{\Gamma(i-1)\Gamma\left(n + 1 + \frac{1}{\consistency-1}\right)}{\Gamma\left(\frac{1}{\consistency - 1} + i\right)\Gamma(n)}} = -\consistency n + \frac{\consistency \Gamma\left(\frac{n\consistency + \consistency - n}{\consistency - 1}\right)}{\Gamma\left(\frac{2\consistency -1}{\consistency - 1}\right)\Gamma(n)} + n - 1.\]
\end{lemma}
\begin{proof}
    We demonstrate first a simpler summation in the form of Equation \eqref{eq:main-aux} below which we argue holds whenever $\consistency > 1$
    \begin{equation}\label{eq:main-aux}
        \sum_{i = 2}^{n}{\frac{\Gamma(i-1)}{\Gamma \left( i + \frac{1}{\consistency - 1} \right)}}  = \frac{(-\consistency n + n - 1)\Gamma(n)}{\Gamma\left(\frac{n\consistency + \consistency - n}{\consistency - 1}\right)} + \frac{\consistency}{\Gamma\left(\frac{2\consistency - 1}{\consistency - 1}\right)}.
    \end{equation}
    Our proof that Equation \ref{eq:main-aux} holds proceeds via induction on $n$.  Consider the base case of $n=2$.  The left-hand side of Eq. \eqref{eq:main-aux} is then \[\frac{\Gamma(1)}{\Gamma\left(2 + \frac{1}{\consistency - 1}\right)} = \frac{1}{\Gamma\left(2 + \frac{1}{\consistency - 1}\right)}.\]
    On the other hand we have that the right-hand side is 
    \begin{align*}
    \frac{(-2\consistency + 1)\Gamma(2)}{\Gamma\left(\frac{2\consistency + \consistency - 2}{\consistency - 1}\right)} + \frac{\consistency}{\Gamma\left(\frac{2\consistency - 1}{\consistency - 1}\right)} &= \frac{(-2\consistency + 1)}{\Gamma\left(\frac{3\consistency - 2}{\consistency - 1}\right)} + \frac{\consistency}{\Gamma\left(\frac{2\consistency - 1}{\consistency - 1}\right)} \\
    &= \frac{1-2\consistency}{\Gamma\left(3 + \frac{1}{\consistency - 1}\right)} + \frac{\consistency}{\Gamma\left(\frac{2\consistency - 1}{\consistency - 1}\right)} \\
    &= \frac{1-2\consistency}{\left(2 + \frac{1}{\consistency - 1}\right)\Gamma\left(2 + \frac{1}{\consistency-1}\right)} + \frac{\consistency}{\Gamma\left(\frac{2\consistency - 1}{\consistency - 1}\right)} \\
    &= \frac{1 - \consistency}{\Gamma\left(2 + \frac{1}{\consistency - 1}\right)} + \frac{\consistency}{\Gamma\left(2 + \frac{1}{\consistency - 1}\right)}\\
    &= \frac{1}{\Gamma\left(2 + \frac{1}{\consistency - 1}\right)},
    \end{align*}
    where the third equality is due to the fact that $\Gamma(z) = (z-1)\Gamma(z-1)$ and the fourth equality is due to the fact that $\consistency > 1$ means that $\consistency \neq 1/2$ (and hence $(1-2\consistency)/(2 + 1/(\consistency - 1)) = 1-2\consistency$).  As such, we have that the base case of our induction holds.  

    We now proceed to the inductive step.  Assume that 
    \[
    \sum_{i = 2}^{n}{\frac{\Gamma(i-1)}{\Gamma\left(i + \frac{1}{\consistency-1}\right)}} = \frac{(-\consistency n + n - 1)\Gamma(n)}{\Gamma\left(\frac{n\consistency + \consistency - n }{\consistency - 1}\right)} + \frac{\consistency}{\Gamma\left(\frac{2\consistency - 1}{\consistency - 1}\right)}
    \] for all $n < n'+1$.  We thus want to show the equation remains true for $n = n'+1$, i.e., we want to show
    \begin{equation}\label{eq:IStep}
        \sum_{i=2}^{n'+1}{\frac{\Gamma(i-1)}{\Gamma\left(i + \frac{1}{\consistency-1}\right)}} = \frac{(-\consistency (n'+1) + (n'+1) - 1)\Gamma(n'+1)}{\Gamma\left(\frac{(n'+1)\consistency + \consistency - (n'+1) }{\consistency - 1}\right)} + \frac{\consistency}{\Gamma\left(\frac{2\consistency - 1}{\consistency - 1}\right)}.
    \end{equation}
    We rewrite the left hand side of Equation \eqref{eq:IStep} as
    \begin{align*}
        \frac{\Gamma(n')}{\Gamma\left(n' + 1 + \frac{1}{\consistency-1}\right)}+\sum_{i = 2}^{n'}{\frac{\Gamma(i-1)}{\Gamma\left(i + \frac{1}{\consistency - 1}\right)}} &= \frac{\Gamma(n')}{\Gamma\left(n' + 1 + \frac{1}{\consistency-1}\right)}+ \frac{(-\consistency n' + n' - 1)\Gamma(n')}{\Gamma\left(\frac{n'\consistency + \consistency - n' }{\consistency - 1}\right)} + \frac{\consistency}{\Gamma\left(\frac{2\consistency - 1}{\consistency - 1}\right)}\\
        &= \frac{\Gamma(n')}{\Gamma\left(\frac{n'(\consistency - 1) + \consistency}{\consistency-1}\right)}+ \frac{(-\consistency n' + n' - 1)\Gamma(n')}{\Gamma\left(\frac{n'(\consistency - 1) + \consistency}{\consistency-1}\right)} + \frac{\consistency}{\Gamma\left(\frac{2\consistency - 1}{\consistency - 1}\right)}\\
        &= \frac{(-\consistency n' +n')\Gamma(n')}{\Gamma\left(\frac{n'(\consistency - 1) + \consistency}{\consistency-1}\right)} + \frac{\consistency}{\Gamma\left(\frac{2\consistency - 1}{\consistency - 1}\right)} \\
        &= \frac{n'(1-\consistency)\Gamma(n')}{\Gamma\left(\frac{n'(\consistency - 1) + \consistency}{\consistency-1}\right)} + \frac{\consistency}{\Gamma\left(\frac{2\consistency - 1}{\consistency - 1}\right)}\\
        &= \frac{(1-\consistency)\Gamma(n'+1)}{\Gamma\left(\frac{n'(\consistency - 1) + \consistency}{\consistency-1}\right)} + \frac{\consistency}{\Gamma\left(\frac{2\consistency - 1}{\consistency - 1}\right)}
    \end{align*}
    where the first equality applies the inductive hypothesis and the last equality applies the identity $\Gamma(z) = (z-1)\Gamma(z-1)$.

    We can similarly rewrite the right hand side of Equation \eqref{eq:IStep} as 
    \begin{align*}
        \frac{(-\consistency (n'+1) + (n'+1) - 1)\Gamma(n'+1)}{\Gamma\left(\frac{(n'+1)\consistency + \consistency - (n'+1) }{\consistency - 1}\right)} + \frac{\consistency}{\Gamma\left(\frac{2\consistency - 1}{\consistency - 1}\right)} &= \frac{((1-\consistency)(n'+1) - 1)\Gamma(n'+1)}{\Gamma\left(\frac{(\consistency - 1)(n'+1)+ \consistency}{\consistency - 1}\right)} + \frac{\consistency}{\Gamma\left(\frac{2\consistency - 1}{\consistency - 1}\right)}\\
        &=\frac{((1-\consistency)(n'+1) - 1)\Gamma(n'+1)}{\Gamma\left(1 + \frac{n'(\consistency - 1)+ \consistency}{\consistency - 1}\right)} + \frac{\consistency}{\Gamma\left(\frac{2\consistency - 1}{\consistency - 1}\right)}\\
        &=\frac{((1-\consistency)(n'+1) - 1)\Gamma(n'+1)}{\frac{n'(\consistency - 1) + \consistency}{\consistency - 1} \cdot \Gamma\left(\frac{n'(\consistency - 1)+ \consistency}{\consistency - 1}\right)} + \frac{\consistency}{\Gamma\left(\frac{2\consistency - 1}{\consistency - 1}\right)}\\
        &=\frac{((1-\consistency)n' - \consistency)\Gamma(n'+1)}{\frac{n'(\consistency - 1) + \consistency}{\consistency - 1} \cdot \Gamma\left(\frac{n'(\consistency - 1)+ \consistency}{\consistency - 1}\right)} + \frac{\consistency}{\Gamma\left(\frac{2\consistency - 1}{\consistency - 1}\right)}\\
        &= \frac{(1-\consistency)\Gamma(n'+1)}{\Gamma\left(\frac{n'(\consistency - 1) + \consistency}{\consistency-1}\right)} + \frac{\consistency}{\Gamma\left(\frac{2\consistency - 1}{\consistency - 1}\right)},
    \end{align*}
    where the third equality applies the identity $\Gamma(z) = (z-1)\Gamma(z-1)$, completing the inductive step as desired.

    Our main summation then follows, essentially, as a direct consequence of Equation \eqref{eq:main-aux}.  Factoring out the common term $\frac{\Gamma\left(n+1+\frac{1}{\consistency - 1}\right)}{\Gamma(n)}$ and applying Equation \eqref{eq:main-aux} we obtain 
    \begin{align*}
        \sum_{i = 2}^{n}{\frac{\Gamma(i-1)\Gamma\left(n + 1 + \frac{1}{\consistency-1}\right)}{\Gamma\left(\frac{1}{\consistency - 1} + i\right)\Gamma(n)}} &= \frac{\Gamma\left(n + 1 + \frac{1}{\consistency-1}\right)}{\Gamma(n)}\sum_{i = 2}^{n}{\frac{\Gamma(i-1)}{\Gamma\left(i + \frac{1}{\consistency-1}\right)}}\\
        &=\frac{\Gamma\left(n + 1 + \frac{1}{\consistency-1}\right)}{\Gamma(n)} \cdot \left(\frac{(-\consistency n + n - 1)\Gamma(n)}{\Gamma\left(\frac{n\consistency + \consistency - n}{\consistency - 1}\right)} + \frac{\consistency}{\Gamma\left(\frac{2\consistency - 1}{\consistency - 1}\right)}\right)\\
        &= -\consistency n + \frac{\consistency \Gamma\left(\frac{n\consistency + \consistency - n}{\consistency - 1}\right)}{\Gamma\left(\frac{2\consistency -1}{\consistency - 1}\right)\Gamma(n)} + n - 1,
    \end{align*} 
    as desired.
\end{proof}

\end{document}